\magnification=1200
\font\Bbb=msbm10
\font\toto=cmbx10 scaled 1200
\font\tbn=cmbx10 scaled 900

\def\ep{\varepsilon}

\def\R{\hbox{\Bbb R}}
\def\S{\hbox{\Bbb S}}
\def\N{\hbox{\Bbb N}}
\def\pa{\partial}
\def\b{\backslash}

{\centerline{\toto On inverse scattering for the multidimensional}}
{\centerline{\toto relativistic Newton equation at high energies}}
\vskip 1cm
{\centerline {\bf A. Jollivet}}
\vskip 4mm
\noindent Laboratoire de Math\'ematiques Jean Leray (UMR 6629), Universit\'e de Nantes, BP 92208, 
F-44322, Nantes cedex 03, France

\noindent e-mail: jollivet@math.univ-nantes.fr

\vskip 1cm

\indent {\tbn Abstract.} 
Consider the Newton equation in the relativistic case (that is the Newton-Einstein equation)
$$
\eqalign
{\dot p = F(x),&\ F(x)=-\nabla V(x),\cr
p={\dot x \over \sqrt{1-{|\dot x|^2 \over c^2}}},&\ \dot p={dp\over dt},\ \dot x={dx\over dt},\ x\in C^1(\R,\R^d),}\eqno{(*)}
$$
$${\rm where\ }V \in C^2(\R^d,\R),\ |\pa^j_x V(x)| \le \beta_{|j|}(1+|x|)^{-(\alpha+|j|)}$$
for $|j| \le 2$ and some $\alpha > 1$.
We give estimates and asymptotics for scattering solutions and scattering data for the equation $(*)$ for the case of small angle
scattering. We show that at high energies the velocity valued component of the scattering operator uniquely determines the X-ray transform 
$PF.$ Applying results on inversion of the X-ray transform $P$ we obtain that for $d\ge 2$ the velocity valued component of the scattering 
operator at high energies uniquely determines $F$. In addition we show that our high energy 
asymptotics found for the configuration valued component of the scattering operator doesn't determine uniquely $F$. 
The results of the present work were obtained in the process of generalizing some results of Novikov [No] to the relativistic case.
\vskip 1cm

{\centerline {\bf 1. Introduction}}
\vskip 2mm

\indent Consider the Newton equation in the relativistic case (that is the Newton-Einstein equation)
$$
\eqalign{\dot p = F(x),&\ F(x)=-\nabla V(x),\cr
p={\dot x \over \sqrt{1-{|\dot x|^2 \over c^2}}},\ \dot p={dp\over dt},&\ \dot x={dx\over dt},\ x\in C^1(\R,\R^d),}
\leqno (1.1)
$$
\vskip -6mm
$${\rm where\ }V \in C^2(\R^d,\R),\ |\pa^j_x V(x)| \le \beta_{|j|}(1+|x|)^{-(\alpha+|j|)}\leqno (1.2)$$
for $|j| \le 2$ and some $\alpha > 1$ (here j is the multiindex $j\in (\N \cup \{0\})^d, |j|= \sum_{n=1}^d j_n$ and $\beta_{|j|}$ are
positive real constants). The equation (1.1) is an equation for $x=x(t)$ and is the equation of motion in $\R^d$ of a relativistic particle of mass $m=1$
and charge $e=1$ in an external electric field described by the scalar potential $V$ (see [E] and, for example, Section 17 of [LL2]). In this equation
$x$ is the position of the particle, $p$ is its impulse, $F$ is the force acting on the particle, $t$ is the time and $c$ is the speed of light.

For the equation (1.1) the energy
$$
E=c^2\sqrt{1+{|p(t)|^2 \over c^2}}+V(x(t))
$$
is an integral of motion.
We denote by $B_c$ the euclidean open ball whose radius is c and whose centre is 0. 
  
\indent   
Yajima [Y] studied in dimension 3 (without loss of generality for the case of dimension $d\ge 2)$ the direct scattering of relativistic particle in 
an external electromagnetic field
described by four vector $(V(x),{\bf A}(x))$ where the scalar potential $V$ and the vector potential ${\bf A}$ are both rapidly decreasing.
We recall the results of Yajima [Y] in our case.
 
\indent Under the conditions (1.2), the following is valid (see [Y]): for any 
$(v_-,x_-)\in B_c\times\R^{d},\ v_-\neq 0,$
the equation (1.1)  has a unique solution $x\in C^2(\R,\R^d)$ such that
$$
{x(t)=v_-t+x_-+y_-(t),}\leqno (1.3)
$$
where $\dot y_-(t)\to 0,\ y_-(t)\to 0,\ {\rm as}\ t\to -\infty;$  in addition for almost any 
$(v_-,x_-)\in B_c\times \R^{d},\ v_-\neq 0,$
$$
{x(t)=v_+t+x_++y_+(t),}\leqno (1.4)
$$
 where $v_+\neq 0,\ |v_+|<c,\ v_+=a(v_-,x_-),\ x_+=b(v_-,x_-),\ \dot y_+(t)\to 0,\ \ y_+(t)\to 0,{\rm\ as\ }t \to +\infty$.

The map $S: B_c\times\R^d \to B_c\times\R^d $ given by the formulas
$$
{v_+=a(v_-,x_-),\ x_+=b(v_-,x_-)}\leqno(1.5)
$$
is called the scattering map for the equation (1.1); in addition, $a(v_-,x_-),$ $b(v_-,x_-)$ are called the scattering data for the equation (1.1).

By ${\cal D}(S)$ we denote the domain of definition of $S$; by ${\cal R}(S)$ we denote the range of $S$ (by definition,
if $(v_-,x_-)\in {\cal D}(S)$, then $v_-\neq 0$ and $a(v_-,x_-)\neq 0$). 

 Under the conditions (1.2), the map $S$  has the following simple properties (see [Y]):
for any $(v,x)\in B_c\times \R^d$, $(v,x)\in {\cal D}(S)$ if and only if $(-v,x)\in {\cal R}(S)$;
${\cal D}(S)$  is an open set of $B_c \times \R^d$ and ${\rm Mes}((B_c \times \R^d) \b {\cal D}(S))=0$ for the Lebesgue
measure on $B_c \times \R^d$  induced by the Lebesgue measure on $\R^d\times\R^d$;
the map $S:{\cal D}(S)\to {\cal R}(S)$ is continuous and preserves the element of volume,
$a(v_-,x_-)^2=v_-^2$.
 
If $V(x)\equiv 0$, then $a(v_-,x_-)=v_-,\ b(v_-,x_-)=x_-,\ (v_-,x_-)\in B_c \times \R^d,\ v_-\neq 0$. Therefore for $a(v_-,x_-),\
b(v_-,x_-)$ we will use the following representation
$$
{\eqalign{
a(v_-,x_-)=&v_-+a_{sc}(v_-,x_-)\cr
b(v_-,x_-)=&x_-+b_{sc}(v_-,x_-)\crcr
}
\hskip 1cm (v_-,x_-)\in {\cal D}(S).}\leqno (1.6)
$$

We will use the fact that, under the conditions (1.2), the map $S$ is uniquely determined by its restriction to ${\cal M}(S)={\cal
D}(S)\cap {\cal M},$ where
$$
{\cal M}=\{(v_-,x_-)\in B_c \times \R^d|v_-\neq 0, v_-x_-=0\}.
$$

Consider 
$$
T\S^{d-1}=\{(\theta,x)|\theta \in \S^{d-1},x\in \R^d,\theta x=0\},
$$
where $\S^{d-1}$ is the unit sphere in $\R^d$.

Consider the X-ray transform $P$ which maps each function $f$ with the properties 
$$
f\in C(\R^d,\R^m),\ |f(x)|=O(|x|^{-\beta}),\ {\rm as\ }|x|\to \infty,{\rm\ for\ some\ }\beta>1
$$
into a function $Pf\in C(T\S^{d-1},\R^m),$ where $Pf$ is defined by
$$
Pf(\theta,x)=\int_{-\infty}^{+\infty}f(t\theta+x)dt,\ (\theta,x)\in T\S^{d-1}.
$$
Concerning the theory of the X-ray transform, the reader is referred to [R], [GGG], [Na] and [No].

Let 
$$
\mu(c,d,\tilde{\beta},\alpha,r_v,r_x,r)
={1\over \sqrt{1+{r_v^2\over 4(c^2-r_v^2)}}}
\times{2^{2\alpha+6}(1+3\tilde{\beta}/c)d^2\sqrt{d}\tilde{\beta}(r_v/\sqrt{2}+1-r)^3 \over r(\alpha-1)(r_v/\sqrt{2}-r)^4
(1+r_x/\sqrt{2})^{\alpha-1}}\leqno(1.7{\rm a})
$$
and let $z=z(c,d,\tilde{\beta},\alpha,r_x,r),$ $z_1=z_1(c,d,\beta_1,\alpha,r_x,r)$ and $z_2=z_2(c,d,\beta_1,\alpha,r_x)$ be defined as the roots of the
following equations
$$
\mu(c,d,\tilde{\beta},\alpha,z,r_x,r)=1,\ z\in]\sqrt{2}r,c[,\leqno(1.7{\rm b})
$$
$$
{z_1\over \sqrt{1-{z_1^2\over c^2}}}-{2^{\alpha+4}\beta_1\sqrt{d}\over \alpha({z_1/\sqrt{2}}-r)(r_x/\sqrt{2}+1)^\alpha}=0
,\ z_1\in ]\sqrt{2}r,c[,\leqno(1.7{\rm c})
$$
$$
{z_2\over \sqrt{1-{z_2^2\over c^2}}}-{8\beta_1 \sqrt{d}\over \alpha (z_2/\sqrt{2})(1+r_x/\sqrt{2})^{\alpha}}=0,\ z_2\in ]0,c[,
\leqno(1.7{\rm d})
$$
where $r_v,$ $r_x$ and $r$ are some nonnegative numbers such that $0< r\le 1,\ r<c/\sqrt{2},$  $\sqrt{2}r<r_v<c,$
and where $\tilde{\beta}=\max(\beta_1,\beta_2).$

The main results of the present work consist in the small angle scattering asymptotics and estimates for the scattering data $a_{sc}$ and $b_{sc}$ 
(and scattering solutions) for the equation (1.1) and in application of these asymptotics and estimates to inverse scattering for the equation (1.1) at 
high energies. Our main results include, in particular, Theorem 1.1 and Proposition 1.1 given below.
\vskip 2mm 
{\bf Theorem 1.1.}
{\it Let the conditions} (1.2) {\it be valid,}\ $\tilde{\beta}=\max(\beta_1,\beta_2),$\ $(\theta,x)\in T\S^{d-1},$
 {\it and let }\ $r$ {\it be a positive constant such that}\ $0<r\le 1,$ $r<c/\sqrt{2}.$ 
{\it Then}
$$
PF(\theta,x)=\lim\limits_{s\to c\atop s<c}{s\over \sqrt{1-{s^2\over c^2}}}a_{sc}(s\theta,x),\leqno(1.8{\rm a})
$$
{\it and, in addition,}
$$
\left|PF(\theta,x)-{s\over \sqrt{1-{s^2\over c^2}}}a_{sc}(s\theta,x)\right|
\le {d^2\tilde{\beta}^22^{2\alpha+5}s(s/\sqrt{2}+1-r)^2\over \sqrt{1+{s^2\over4(c^2-s^2)}}\alpha (\alpha-1)(s/\sqrt{2}-r)^4
(1+|x|/\sqrt{2})^{2\alpha-1}}\leqno(1.8{\rm b})
$$
{\it for\ }$s<c,\ s>z(c,d,\tilde{\beta},\alpha,|x|,r),\ s\ge z_1(c,d,\beta_1,\alpha,|x|,r);$
$$
\int_{-\infty}^0\!\int_{-\infty}^{\tau} F(s\theta+x)dsd\tau-\int_0^{+\infty}\!\int^{+\infty}_{\tau} F(s\theta+x)dsd\tau
+PV(\theta,x)\theta=\lim\limits_{s\to c\atop s<c}{s^2\over \sqrt{1-{s^2\over c^2}}}b_{sc}(s\theta,x).\leqno(1.9{\rm a})
$$
{\it and, in addition,}
$$
\leqalignno{
\left|{b_{sc}(s\theta,x)\over \sqrt{1-{s^2\over c^2}}}
-{1\over c^2} PV(\theta,x)\theta\right.&\left.+{1\over s^2}\int_0^{+\infty}\!\!\!\int_{\tau}^{+\infty}F(u\theta+x)dud\tau
-{1\over s^2}\int^0_{-\infty}\!\!\int^{\tau}_{-\infty}F(u\theta+x)dud\tau\right|\cr
\le&\sqrt{1-{s^2\over c^2}}\left[C+
{d^3\sqrt{d}(\beta_2+3{\beta_1\beta_2\over c})\beta_12^{3\alpha+8}({s\over \sqrt{2}}+1-r)^3\over
(1-{3\over 4}{s^2\over c^2})\alpha (\alpha-1)^2({s\over \sqrt{2}}-r)^6(1+{|x|\over \sqrt{2}})^{2\alpha-2}}\right]&(1.9{\rm b}) 
}
$$
{\it for\ }$\ s<c,\ s>z(c,d,\tilde{\beta},\alpha,|x|,r),\ s\ge \max(z_1(c,d,\beta_1,\alpha,|x|,r),z_2(c,d,\beta_1,\alpha,|x|))$ {\it and some constant} $C=
C(c,d,\beta_0,\beta_1,\alpha,|x|)$ {\it which can be given explicitly.}
\vskip 2mm

\vskip 2mm
Consider the vector-function $w$ of $(\theta,x)$ arising in the left-hand side of (1.9a) :
$$
\eqalign{
w(V,\theta,x)=&\int_{-\infty}^0\!\int_{-\infty}^{\tau} F(s\theta+x)dsd\tau-\int_0^{+\infty}\!\int^{+\infty}_{\tau} F(s\theta+x)dsd\tau\cr
&+PV(\theta,x)\theta,\ (\theta,x)\in T\S^{d-1} .
}
$$
\vskip 2mm
{\bf Proposition 1.1.}
{\it The vector} $w$ {\it as a function of potential} $V$ {\it satisfying the conditions} (1.2) {\it and of} $(\theta,x)\in T\S^{d-1}$ 
{\it has the following simple properties:}

1.{\it\ under the conditions} (1.2), {\it for any potential} $V$ {\it the vector} $w(V,\theta,x)$ {\it is orthogonal to} 

$\theta$,
 
2.{\it\ there exists a potential} $V$ {\it which satisfies the conditions} (1.2) {\it and for which}  $w(V,\theta,x)$ 

{\it isn't null for all} $(\theta,x)\in T\S^{d-1},$

3.{\it\ for any spherical symmetric potential} $V$ {\it satisfying the conditions} (1.2) {\it we have} 

$w(V,\theta,x)=0$ {\it for all}  $(\theta,x)\in T\S^{d-1}$ .

\vskip 2mm
From (1.8a) and inversion formulas for the X-ray transform for $d\ge 2$ (see [R], [GGG], [Na], [No]) it follows that $a_{sc}$ determines 
uniquely $F$ at high energies. Moreover for $d\ge 2$ methods of reconstruction of $f$ from $Pf$ (see [R], [GGG], [Na], [No])
permit to reconstruct $F$ from the velocity valued component $a$ of the scattering map at high energies. The formula 
(1.9a) and the item 3 of Proposition 1.1 show that the first term of the asymptotics of $b_{sc}$ doesn't determine
uniquely the potential $V$ or the force $F$. The item 2 of Proposition 1.1 ensures us that the asymptotics which was found for $b_{sc}$ is 
nontrivial. Note that F. Nicoleau paid our attention to the fact that, in addition of the item 3 of Proposition 1.1, $w(V,\theta,x), (\theta,x)\in
T\S^{d-1},$ uniquely determines $V$ satisfying (1.2) modulo spherical symmetric potentials. 

Inverse scattering for the classical multidimensional Newton equation was first studied by Novikov [No] (the
existence of the scattering states, asymptotic completness and scattering map for the classical Newton equation was 
studied by Simon [S]). Novikov proved two formulas which link scattering data 
at high energies to the X-ray transform of $F$ and $V$. These formulas are generalized to the relativistic case by the formulas 
(1.8a) and (1.9a) of Theorem 1.1. Then applying results on inversion of the X-ray transform, Novikov obtains that at high energies the
velocity valued component of the scattering data determines uniquely the X-ray transform of $F$ whereas the configuration valued component of the scattering 
operator determines uniquely the X-ray transform of $V$. Note that in the relativistic case (due to the formula (1.9a) and Proposition 1.1)
the asymptotics of $b_{sc}$ doesn't determine uniquely $F$. We follow Novikov's framework [No] to obtain our results.
Note also that for the classical multidimensional Newton equation in a bounded open strictly convex domain an inverse boundary value
problem at high energies was first studied in [GN].  

Further our paper is organized as follows. In Section 2 we transform the differential equation (1.1) with initial conditions (1.3) in
an integral equation which takes the form $y_-=A_{v_-,x_-}(y_-).$ Then we study $A_{v_-,x_-}$ on a
suitable space and we give estimates and contraction estimates about $A_{v_-,x_-}$ (Lemmas 2.1, 2.2, 2.3). In Section 3 we give estimates and asymptotics for the deflection $y_-(t)$ from (1.3) and for  
scattering data $a_{sc}(v_-,x_-),b_{sc}(v_-,x_-)$ from (1.6) (Theorem 3.1 and Theorem 3.2). From these estimates and asymptotics the 
 formulas (1.8a) and (1.9a) will follow when the parameters $c,\beta_m,\ \alpha,\ d,\ {\hat v}_-,\ x_-$ are fixed and $|v_-|$ increases (where 
$\beta_{|j|},\ \alpha,\ d$ are constants from (1.2),
$\beta_m=\max(\beta_0,\beta_1,\beta_2);$ ${\hat v}_-=v_-/|v_-|).$ In these cases $\sup_{t\in \R}|\theta(t)|$ decreases, where $\theta(t)$ denotes the
angle between the vectors ${\dot x}(t)=v_-+{\dot y}_-(t)$ and $v_-,$ and we deal with small angle scattering. 
Note that, under the conditions of Theorem 3.1, without additional assumptions, there is the estimate $\sup_{t\in \R}|\theta(t)|<
{1\over4}\pi$ and we deal with rather small angle scattering (concerning the term ``small angle scattering" see [No] and Section 20 of
[LL1]).
 Theorem 1.1 follows from Theorem 3.1 and Theorem 3.2. Section 4, Section 5 and Section 6 are devoted to Proofs of our Theorems and
 Lemmas. 

\vfill\eject

{\centerline {\bf 2. A contraction map}}
\vskip 4mm
Let us transform the differential equation (1.1) in an integral equation. Consider the function $g:\R^d\to B_c$ defined by
$$g(x)={x\over \sqrt{1+{|x|^2\over c^2}}}$$
where $x\in \R^d$. One can see that $g$ has, in particular, the following simple properties :
$${|g(x)-g(y)|\le \sqrt{d} |x-y|,\ {\rm for\ } x,y\in\R^d,}\leqno(2.1)$$
$g$ is an infinitely smooth diffeomorphism between $\R^d$ and $B_c$, and its inverse is given by 
$$\gamma(x)={x\over \sqrt{1-{|x|^2\over c^2}}},\ x\in B_c.$$

Now, if $x$ satisfies the differential equation (1.1) and the initial conditions (1.3), then $x$ satisfies the integral equation
$$
x(t)=v_-t+x_-+\int\limits_{-\infty}^t\!\left[g\left(\gamma(v_-)+\int\limits_{-\infty}^\tau\! F(x(s))ds\right)-v_-\right]d\tau,
\leqno(2.2)
$$
where $F(x)=-\nabla V(x),\ v_-\in B_c\b\{0\}.$

For $y_-(t)$ this equation takes the form
$$
y_-(t)=A_{v_-,x_-}(y_-)(t),
\leqno(2.3)
$$
${\rm where\ }
A_{v_-,x_-}(f)(t)=\int\limits_{-\infty}^t\left[g(\gamma(v_-)+\int\limits_{-\infty}^\tau F(v_-s+x_-+f(s))ds)-v_-\right]d\tau,\ v_-\in
B_c\b\{0\}.
$

From (2.2), (1.2), (2.1) (applied on  ``$x$"$=\gamma (v_-)+\int_{-\infty}^{\tau}F(x(s))ds$ and ``$y$"$=\gamma(v_-)$) and 
$y_-(t)\in C(\R,\R^d),\ y_-(t)\to 0,$ as $t\to -\infty$, it follows, in particular, that
$$
y_-(t)\in C^1(\R,\R^d)\ {\rm and\ } |\dot y_-(t)|=O(|t|^{-\alpha}),\ |y_-(t)|=O(|t|^{-\alpha+1}),\ {\rm as\ }t \to -\infty, \leqno(2.4)
$$
where $v_-\in B_c\b\{ 0\}$ and $x_-$ are fixed.

Consider the complete metric space
$$
\eqalign{
M_{T,r}=&\{f\in C^1(]-\infty,T],\R^d)|\ \|f\|_T\le r\},\cr
{\rm where\ }\|f\|_T=&\max\left(\sup\limits_{t\in ]-\infty,T]}|\dot f(t)|,\sup\limits_{t\in ]-\infty,T]}|f(t)-t\dot f(t)|\right) \crcr
}\leqno(2.5)
$$
(where for $T=+\infty$ we understand $]-\infty, T]$ as $]-\infty,+\infty[$). From (2.4)
it follows that, at fixed $T<+\infty$,
$$
y_-(t)\in M_{T,r}  {\rm \ for\ some\ }r\ {\rm depending\ on}\ y_-(t)\ {\rm and}\ T.\leqno(2.6)
$$

\vskip 2mm

{\bf Lemma 2.1.} {\it Under the conditions} (1.2), {\it the following is valid: if} $f\in M_{T,r},\ 0< r\le 1,\ r<c/\sqrt{2},$
$|v_-|<c,$ $|v_-|
\ge z_1(c,d,\beta_1,\alpha,|x_-|,r),\ v_-x_-=0,$ {\it then}
$$
\leqalignno{
\| A_{v_-,x_-}(f)\|_T\le& \rho_T(c,d,\beta_1,\alpha,|v_-|,|x_-|,r)\cr
=&{1\over \sqrt{1+|v_-|^2/(4(c^2-|v_-|^2))}}
{2^{\alpha+1}d\beta_1(|v_-|/\sqrt{2}+1-r) \over (\alpha-1)(|v_-|/\sqrt{2}-r)^2(1+|x_-|/\sqrt{2}-(|v_-|/\sqrt{2}-r)T)^{\alpha-1}}
&(2.7{\rm a})\cr
}
$$
{\it for} $T\le 0$,
$$
\leqalignno{
\| A_{v_-,x_-}(f)\|_T\le&\rho(c,d,\beta_1,\alpha,|v_-|,|x_-|,r)\cr
=&{1\over \sqrt{1+|v_-|^2/(4(c^2-|v_-|^2))}}
{2^{\alpha+2}d\beta_1(|v_-|/\sqrt{2}+1-r) \over (\alpha-1)(|v_-|/\sqrt{2}-r)^2(1+|x_-|/\sqrt{2})^{\alpha-1}}
&(2.7{\rm b})\cr
}
$$
{\it for} $T\le +\infty$;

{\it if} $f_1,f_2\in M_{T,r},\ 0< r\le 1,\ r<c/\sqrt{2},\ |v_-|<c,$ $|v_-|\ge z_1(c,d,\beta_1,\alpha,|x_-|,r),\ v_-x_-=0,$ {\it then}
$$
\| A_{v_-,x_-}(f_2)-A_{v_-,x_-}(f_1)\|_T\le \lambda_T(c,d,\beta_2,\alpha,|v_-|,|x_-|,r)\|f_2-f_1\|_T,
\leqno(2.8{\rm a})
$$
$$
\eqalign{\lambda_T(c,d,\beta_2,\alpha,|v_-|,|x_-|,r)=&{1\over \sqrt{1+|v_-|^2/(4(c^2-|v_-|^2))}(\alpha-1)}\cr
&\times{2^{\alpha+2}d\sqrt{d}\beta_2(|v_-|/\sqrt{2}+1-r)^2 \over (|v_-|/\sqrt{2}-r)^3(1+|x_-|/\sqrt{2}-(|v_-|/\sqrt{2}-r)T)^{\alpha-1}}\cr
}
$$
{\it for} $T\le 0$,
$$
\| A_{v_-,x_-}(f_2)-A_{v_-,x_-}(f_1)\|_T\le \lambda(c,d,\beta_1,\beta_2,\alpha,|v_-|,|x_-|,r)\|f_2-f_1\|_T,
\leqno(2.8{\rm b})
$$
$$
\eqalign{
\lambda(c,d,\beta_1,\beta_2,\alpha,|v_-|,|x_-|,r)=&{1\over \sqrt{1+|v_-|^2/(4(c^2-|v_-|^2))}}\cr
&\times{2^{2\alpha+6}(\beta_2+3\beta_1\beta_2/c)d^2\sqrt{d}(|v_-|/\sqrt{2}+1-r)^3 \over (\alpha-1)(|v_-|/\sqrt{2}-r)^4(1+|x_-|/\sqrt{2}
)^{\alpha-1}}\cr
}
$$
{\it for} $T\le +\infty.$
\vskip 4mm
Note that 
$$
\max\left({\rho_T(c,d,\beta_1,\alpha,|v_-|,|x_-|,r)\over r}, \lambda_T(c,d,\beta_2,\alpha,|v_-|,|x_-|,r) \right)\ \hskip 4cm
$$
\vskip -8mm
$$
\leqalignno{
\le &\mu_T(c,d,\tilde{\beta},\alpha,|v_-|,|x_-|,r)&(2.9{\rm a})\cr
 =&{1\over \sqrt{1+|v_-|^2/(4(c^2-|v_-|^2))}}\cr
&\times{2^{\alpha+2}d\sqrt{d}\tilde{\beta}(|v_-|/\sqrt{2}+1-r)^2 \over r(\alpha-1)(|v_-|/\sqrt{2}-r)^3(1+|x_-|/\sqrt{2}-
(|v_-|/\sqrt{2}-r)T)^{\alpha-1}}\cr
}
$$
for $T\le 0$,
$$
\max\left({\rho(c,d,\beta_1,\alpha,|v_-|,|x_-|,r)\over r}, \lambda(c,d,\beta_1,\beta_2,\alpha,|v_-|,|x_-|,r)\right)\ \hskip 4cm
$$
\vskip -8mm
$$
\leqalignno{\le &\mu(c,d,\tilde{\beta},\alpha,|v_-|,|x_-|,r)&(2.9{\rm b})\cr
=&{1\over \sqrt{1+|v_-|^2/(4(c^2-|v_-|^2))}}\cr
&\times{2^{2\alpha+6}(1+3\tilde{\beta}/c)d^2\sqrt{d}\tilde{\beta}(|v_-|/\sqrt{2}+1-r)^3 \over r(\alpha-1)(|v_-|/\sqrt{2}-r)^4
(1+|x_-|/\sqrt{2})^{\alpha-1}}\cr
}
$$
for $T\le +\infty$, where $\tilde{\beta}=\max(\beta_1,\beta_2),\ 0< r\le 1,\ r<c/\sqrt{2},\ |v_-|<c,$ $|v_-|
\ge z_1(c,d,\beta_1,\alpha,|x_-|,r),$
$\ v_-x_-=0.$

From Lemma 2.1 and the estimates (2.9) we obtain the following result.
\vskip 2mm
{\bf Corollary 2.1.} {\it Under the conditions} (1.2),$\ 0< r\le 1,$ $r<c/\sqrt{2},$ $|v_-|<c,$ $|v_-|\ge$

\noindent $z_1(c,d,\beta_1,\alpha,|x_-|,r),\ v_-x_-=0,$ {\it the following result is valid:}

{\it if} $\mu_T(c,d,\tilde{\beta},\alpha,|v_-|,|x_-|,r)<1$, {\it then} $A_{v_-,x_-}$ {\it is a contraction map in }$M_{T,r}$ 
{\it for} $T\le 0;$ 

{\it if} $\mu(c,d,\tilde{\beta},\alpha,|v_-|,|x_-|,r)<1$, {\it then} $A_{v_-,x_-}$ {\it is a contraction map in }$M_{T,r}$ 
{\it for} $T\le +\infty$.
\vskip 4mm
Taking into account (2.6) and using Lemma 2.1, Corollary 2.1 and the lemma about the contraction maps we will study the solution $y_-(t)$ 
of the equation (2.3) in $M_{T,r}$.

We will use also the following results.

\vskip 2mm

{\bf Lemma 2.2.} {\it Under the conditions} (1.2), $f\in  M_{T,r},\ 0< r\le 1,\ r<c/\sqrt{2},$ $|v_-|<c,$ 
$|v_-|\ge z_1(c,d,\beta_1,\alpha,|x_-|,r),\ v_-x_-=0,$ {\it the following is valid:}
$$
\leqalignno{
|\dot A_{v_-,x_-}(f)(t)|\le&\zeta_-(c,d,\beta_1,\alpha,|v_-|,|x_-|,r,t)\cr
=&{1\over \sqrt{1+|v_-|^2/(4(c^2-|v_-|^2))}}&(2.10)\cr
&\times{d\beta_1 2^{\alpha+1}\over \alpha (|v_-|/\sqrt{2}-r)(1+|x_-|/\sqrt{2}-(|v_-|/\sqrt{2}-r)t)^\alpha},\cr
|A_{v_-,x_-}(f)(t)|\le&\xi_-(c,d,\beta_1,\alpha,|v_-|,|x_-|,r,t)\cr
=&{1\over \sqrt{1+|v_-|^2/(4(c^2-|v_-|^2))}}&(2.11)\cr
&\times{d\beta_1 2^{\alpha+1}\over (\alpha-1) \alpha (|v_-|/\sqrt{2}-r)^2(1+|x_-|/\sqrt{2}-(|v_-|/\sqrt{2}-r)t)^{\alpha-1}},\crcr
}
$$
{\it for} $t\le T,\ T\le 0;$ 
$$
A_{v_-,x_-}(f)(t)=k_{v_-,x_-}(f)t+l_{v_-,x_-}(f)+H_{v_-,x_-}(f)(t),\leqno(2.12)
$$
{\it where}
$$
\leqalignno{
k_{v_-,x_-}(f)=&g(\gamma (v_-)+\int_{-\infty}^{+\infty}\! F(v_-s+x_-+f(s))\,ds)-v_-,&(2.13{\rm a})\cr
l_{v_-,x_-}(f)=&\int_{-\infty}^0\!\left[g(\gamma (v_-)+\int_{-\infty}^{\tau}\! F(v_-s+x_-+f(s))\,ds)-v_-\right]\,d\tau
&(2.13{\rm b})\cr
                   &+\int^{+\infty}_0\!\left[g(\gamma (v_-)+\int_{-\infty}^{\tau}\! F(v_-s+x_-+f(s))\,ds)\right.\cr
                   &\left.-g(\gamma (v_-)+\int_{-\infty}^{+\infty}\! F(v_-s+x_-+f(s))\,ds)\right]\,d\tau,\cr
|k_{v_-,x_-}(f)|\le&2\zeta_-(c,d,\beta_1,\alpha,|v_-|,|x_-|,r,0),&(2.14{\rm a})\cr
|l_{v_-,x_-}(f)|\le&2\xi_-(c,d,\beta_1,\alpha,|v_-|,|x_-|,r,0),&(2.14{\rm b})\cr\cr
}
$$
\vskip -8mm
$$
\leqalignno{
|\dot H_{v_-,x_-}(f)(t)|\le& \zeta_+(c,d,\beta_1,\alpha,|v_-|,|x_-|,r,t)&(2.15)\cr
                          =&{1\over \sqrt{1+|v_-|^2/(4(c^2-|v_-|^2))}\alpha (|v_-|/\sqrt{2}-r)}\cr
			   &\times{d\beta_1 2^{\alpha+1}\over (1+|x_-|/\sqrt{2}+(|v_-|/\sqrt{2}-r)t)^\alpha},\cr
|H_{v_-,x_-}(f)(t)|\le& \xi_+(c,d,\beta_1,\alpha,|v_-|,|x_-|,r,t)&(2.16)\cr
                     =&{1\over \sqrt{1+|v_-|^2/(4(c^2-|v_-|^2))}\alpha (\alpha-1) (|v_-|/\sqrt{2}-r)^2}\cr
                      &\times{d\beta_1 2^{\alpha+1}\over (1+|x_-|/\sqrt{2}+(|v_-|/\sqrt{2}-r)t)^{\alpha-1}},\crcr
}		      
$$
{\it for} $T=+\infty,\ t\ge 0.$

\vskip 2mm

{\bf Lemma 2.3.} {\it Let the conditions} (1.2) {\it be valid,} $y_-(t)\in  M_{T,r}$ {\it be a solution of}\  (2.3), $T=+\infty,
\ 0< r\le 1,\ r<c/\sqrt{2},$ $|v_-|<c,$ $|v_-|\ge z_1(c,d,\beta_1,\alpha,|x_-|,r),\ v_-x_-=0,$ {\it then}
$$
\leqalignno{
|k_{v_-,x_-}(y_-)-k_{v_-,x_-}(0)|\le & \ep_a'(c,d,\beta_1,\beta_2,\alpha,|v_-|,|x_-|,r)\cr
=&{d\sqrt{d} \beta_2 2^{\alpha+3}(|v_-|/\sqrt{2}+1-r)\over \alpha (|v_-|/\sqrt{2}-r)^2(1+|x_-|/\sqrt{2})^\alpha}&(2.17{\rm a})\cr
& \times {\rho(c,d,\beta_1,\alpha,|v_-|,|x_-|,r) \over \sqrt{1+|v_-|^2/(4(c^2-|v_-|^2))}},\cr
\left|{k_{v_-,x_-}(y_-)\over\sqrt{1-{|v_-|^2\over c^2}}}-\int_{-\infty}^{+\infty}F(x_-+v_-s)\,ds\right|\le&
\ep_a(c,d,\beta_1,\beta_2,\alpha,|v_-|,|x_-|,r)\cr
=&{d\beta_2 2^{\alpha+3}(1+|v_-|/\sqrt{2}-r)
\rho(c,d,\beta_1,\alpha,|v_-|,|x_-|,r)\over \alpha (|v_-|/\sqrt{2}-r)^2(1+|x_-|/\sqrt{2})^\alpha},&(2.17{\rm b})\cr
|l_{v_-,x_-}(y_-)-l_{v_-,x_-}(0)|\le & \ep_b(c,d,\beta_1,\beta_2,\alpha,|v_-|,|x_-|,r)\cr
=&{d^2\sqrt{d}(\beta_2+3\beta_1\beta_2/c) 2^{2\alpha+6}(|v_-|/\sqrt{2}+1-r)^2\over \alpha (\alpha-1) (|v_-|/\sqrt{2}-r)^4 (1+|x_-|/\sqrt{2})^{\alpha-1}}
&(2.17{\rm c})\cr
& \times{\rho(c,d,\beta_1,\alpha,|v_-|,|x_-|,r) \over \sqrt{1+|v_-|^2/(4(c^2-|v_-|^2))}}.\crcr
}
$$
\vskip 2mm
Proofs of Lemmas 2.1, 2.2, 2.3 are given in Section 5.
\vskip 1cm

{\centerline {\bf 3. Small angle scattering}}
\vskip 4mm
Under the conditions (1.2), for any $(v_-,x_-)\in B_c\times \R^d,\ v_-\neq 0,$ the equation (1.1) has a unique solution $x \in
C^2(\R,\R^d)$ with the initial conditions (1.3). Consider the function $ y_-(t)$ from (1.3). This function describes deflection from 
free motion.

Using Corollary 2.1 the lemma about contraction maps, and Lemmas 2.2 and 2.3 we obtain the following result.
\vskip 2mm
{\bf Theorem 3.1.}
{\it Let the conditions} (1.2) {\it be valid},
$\mu(c,d,\tilde{\beta},\alpha,|v_-|,|x_-|,r)<1$, $\tilde{\beta}=\max(\beta_1,\beta_2),\ 0< r\le 1,\ r<c/\sqrt{2},$ $|v_-|<c,$ $|v_-|\ge
z_1(c,d,\beta_1,\alpha,|x_-|,r),\ v_-x_-=0$ . {\it Then the deflection} $y_-(t)$ {\it has the following
properties:}
$$y_-\in M_{T,r},\ T=+\infty;\leqno(3.1)$$
$$
\leqalignno{
|{\dot y}_-(t)|\le&\zeta_-(c,d,\beta_1,\alpha,|v_-|,|x_-|,r,t),&(3.2)\cr
|y_-(t)|\le& \xi_-(c,d,\beta_1,\alpha,|v_-|,|x_-|,r,t)\ \ {\it for}\ \ t\le 0;
&(3.3)\cr
y_-(t)=&a_{sc}(v_-,x_-)t + b_{sc}(v_-,x_-) + h(v_-,x_-,t),&(3.4)\crcr
}
$$
{\it where}
$$
\left|a_{sc}(v_-,x_-)-\left[{\gamma(v_-) +\int_{-\infty}^{+\infty}
F(v_-s + x_-)ds \over \sqrt{1+{|\gamma(v_-) +\int_{-\infty}^{+\infty}F( v_-s + x_-)ds|^2\over c^2}}}-v_-\right]\right|
$$
$$
\le \ep_a'(c,d,\beta_1,\beta_2,\alpha,|v_-|,|x_-|,r),\leqno(3.5{\rm a})
$$
$$
\leqalignno{
\left|{a_{sc}(v_-,x_-)\over \sqrt{1-{|v_-|^2\over c^2}}}\right.-\left.\int_{-\infty}^{+\infty}F(v_-s + x_-)ds\right|
\le &\ep_a(c,d,\beta_1,\beta_2,\alpha,|v_-|,|x_-|,r)&(3.5{\rm b})\cr
|b_{sc}(v_-,x_-)-l_{v_-,x_-}(0)|\le& \ep_b(c,d,\beta_1,\beta_2,\alpha,|v_-|,|x_-|,r),&(3.5{\rm c})
}
$$
$$
\leqalignno{
|a_{sc}(v_-,x_-)|\le& 2\zeta_-(c,d,\beta_1,\alpha,|v_-|,|x_-|,r,0),&(3.6{\rm a})\cr
|b_{sc}(v_-,x_-)|\le& 2\xi_-(c,d,\beta_1,\alpha,|v_-|,|x_-|,r,0),& (3.6{\rm b})\cr
|{\dot h}(v_-,x_-,t)|\le& \zeta_+(c,d,\beta_1,\alpha,|v_-|,|x_-|,r,t),&(3.7)\cr
|h(v_-,x_-,t)|\le& \xi_+(c,d,\beta_1,\alpha,|v_-|,|x_-|,r,t)&(3.8)\cr
}
$$
{\it for} $t\ge 0$, {\it where} $l_{v_-,x_-}(0)$ {\it (resp.} $\ep_a',$ $\ep_a,$ $\ep_b,$ $\zeta_-,$ $\zeta_+,$ $\xi_-$ {\it and} $\xi_+$)
{\it is defined in} (2.13b) ({\it resp.} (2.17a), (2.17b), (2.17c), (2.10), (2.15), (2.11) {\it and} (2.16)).

\vskip 4mm
We will use the following observations.

(I) Let $\ 0< r\le 1, r<c/\sqrt{2},\ 0\le u$
$$
{s_1\over \sqrt{1-{s_1^2\over c^2}}}-{2^{\alpha+4}\beta_1\sqrt{d}\over \alpha({s_1/\sqrt{2}}-r)(u/\sqrt{2}+1)^\alpha}
>
{s_2\over \sqrt{1-{s_2^2\over c^2}}}-{2^{\alpha+4}\beta_1\sqrt{d}\over \alpha({s_2/\sqrt{2}}-r)(u/\sqrt{2}+1)^\alpha}
$$
for $\sqrt{2}r <s_2<s_1<c$.

(II) Let $\ 0< r\le 1, r<c/\sqrt{2},\ u\in ]\sqrt{2}r,c[,$
$$
{u\over \sqrt{1-{u^2\over c^2}}}-{2^{\alpha+4}\beta_1\sqrt{d}\over \alpha({u/\sqrt{2}}-r)(s_1/\sqrt{2}+1)^\alpha}
>
{u\over \sqrt{1-{u^2\over c^2}}}-{2^{\alpha+4}\beta_1\sqrt{d}\over \alpha({u/\sqrt{2}}-r)(s_2/\sqrt{2}+1)^\alpha}
$$
for $0\le s_2<s_1$. 

(III) Let $0< r\le 1,\ r<c/\sqrt{2},$ $x$ some real nonnegative number, $\tilde{\beta}=\max(\beta_1,\beta_2)$ and $\sqrt{2}r<s<c$ then
$$
\mu(c,d,\tilde{\beta},\alpha,s,|x|,r)<1\Leftrightarrow s>z(c,d,\tilde{\beta},\alpha,|x|,r).
$$
Observations (I) and (II) imply that  $z_1(c,d,\beta_1,\alpha,s_2,r)>z_1(c,d,\beta_1,\alpha,s_1,r)$ for $\sqrt{2}r<s_2<s_1<c$ when $c,\ \beta_1,\ \alpha,
\  d,\ r$ are fixed. 

Theorem 3.1 gives, in particular, estimates for the scattering process and asymptotics for the velocity valued component of
the scattering map when $c,\ \beta_1,\ \beta_2,\ \alpha,\ d,\ \hat v_-,\ x_-$ are fixed (where $\hat v_-=v_-/|v_-|$) and $|v_-|$ increases or, 
e.g., $c,\ \beta_1,\ \beta_2,\ \alpha,\  d,\ v_-,\ \hat x_-$ are fixed and $|x_-|$ increases. In these cases $\sup_{t\in \R}|\theta(t)|$ 
decreases, where $\theta(t)$ denotes the angle between the vectors $\dot x(t)=v_-+\dot y_-(t)$ and $v_-$, and we deal with small angle 
scattering. Note that already under the conditions of Theorem 3.1, without additional assumptions, there is the estimate 
$\sup_{t\in \R}|\theta(t)|<{1\over 4}\pi$ and we deal with a rather small angle scattering. 
Theorem 3.1 with (3.5c) will give the asymptotics of the configuration valued component $b(v_-,x_-)$ of the scattering map if we can study the 
asymptotics of $l_{v_-,x_-}(0)$. This is the subject of Theorem 3.2.

\vskip 2mm

{\bf Theorem 3.2.} {\it Let} $c,\ d,\ \beta_0,\ \beta_1,\ \alpha,\ |x|$ {\it be fixed. Then there exists  a constant} 
$C_{c,d,\beta_0,\beta_1,\alpha,|x|}$ {\it \ such that\ }
$$
\leqalignno{
\left|{l_{v,x}(0)\over \sqrt{1-{|v|^2\over c^2}}}
-{1\over c^2} PV({\hat v},x){\hat v}\right.&\left.+{1\over|v|^2}\int_0^{+\infty}\!\!\!\int_{\tau}^{+\infty}F(u{\hat v}+x)dud\tau
-{1\over|v|^2}\int^0_{-\infty}\!\!\int^{\tau}_{-\infty}F(u{\hat v}+x)dud\tau\right|\cr
\le&C_{c,d,\beta_0,\beta_1,\alpha,|x|}\sqrt{1-{|v|^2\over c^2}}&(3.9) 
}
$$
{\it for any }$\ v\in B_c,$ $|v|\ge z_2(c,d,\beta_1,\alpha,|x|),$ $vx=0,$ {\it and where }
${\hat v}=v/|v|.$
\vskip 2mm
The proof of Theorem 3.2 is given in Section 6. Using this proof one can compute $C_{c,d,\beta_0,\beta_1,\alpha,|x|}$ explicitly. 
\vskip 1cm

{\centerline{\bf 4. Preliminaries for the main proofs}}
\vskip 4mm
{\it 4.1 Inequalities for $F$.}

{\bf Lemma 4.1.} {\it Under the conditions} (1.2), {\it the following estimates are valid:}
$$
\leqalignno{
|F(x)|=&\left( \sum_{j=1}^d|{\pa\over \pa{x_j}} V(x)|^2\right)^{1\over 2} \le \beta_1\sqrt{d}(1+|x|)^{-(\alpha+1)}\ {\it for\ }x\in\R^d,
&(4.1)\cr
|F(x)-F(y)|\le &\beta_2d \sup\limits_{\ep\in [0,1]}(1+|\ep x+(1-\ep) y|)^{-(\alpha+2)}|x-y|,{\it\ for\ } x,\ y\in \R^d.&(4.2)
}
$$
\vskip 2mm
Lemma 4.1 follows directly from the formula $F(x)=-\nabla V(x)$ and the conditions (1.2).
\vskip 2mm
{\it 4.2 Infinitely smooth function $g:\R^d\to B_c$.}

{\bf Lemma 4.2.}  {\it The following estimates hold:}
$$
\leqalignno{
|\nabla g_i(x)|^2\le& {1\over{1+{|x|^2\over c^2}}}\ {\it for\ }x\in\R^d,\ i=1..d,&(4.3)\cr
|g(x)-g(y)|\le& \sqrt{d} \sup\limits_{\ep\in [0,1]}{1\over\sqrt{1+{|\ep x+(1-\ep) y|^2\over c^2}}}|x-y|,{\it\ for\ } x,\ y\in \R^d,
&(4.4)\cr
|\nabla g_j(x)-\nabla g_j(y)|\le& {3 \sqrt{d}\over c} \sup\limits_{\ep\in [0,1]}{1\over1+{|\ep x+(1-\ep) y|^2\over c^2}} |x-y|,{\it\ for\ } x,
\ y\in \R^d.
&(4.5)\crcr
}
$$
{\it where $g=(g_1,..,g_d).$}
\vskip 2mm 
Lemma 4.2 follows from straighforward calculations.
\vskip 2mm 
{\bf Remark 4.1.} Using the growth properties of $g(p)$ with respect to $|p|$ and following Novikov's framework [No], we will easily generalize some of 
the results of [No] to the relativistic case. Note that ${1\over1+ |p|^2/c^2}\to 0$ when  $p\in \R^d,\ |p|\to +\infty.$

\vskip 2mm
{\it 4.3 Some estimates of integrals.}

We will use the following estimates.  
For $a>0,\ b>0,\ \beta>1,$
$$
\leqalignno{
\int\limits_{-\infty}^t(a+b|s|)^{-\beta}\,\,ds=&{1\over (\beta -1) b (a-bt)^{\beta-1}},{\rm\ for\ }t\le 0,&(4.6)\cr
\int\limits_{-\infty}^t(a+b|s|)^{-\beta}\,\,ds\le&{2\over (\beta -1) b a^{\beta-1}},{\rm\ for\ }t\ge 0.&(4.7)\crcr
}
$$

For $a>0,\ b>0,\ \beta >2,$
$$
\leqalignno{
\int\limits_{-\infty}^t\!\int\limits_{-\infty}^\tau(a+b|s|)^{-\beta}\,\,ds\,d\tau=&{1\over (\beta -2)(\beta -1) b^2 (a-bt)^{\beta-2}},
{\rm\ for\ }t\le 0,&(4.8)\cr
\int\limits_{0}^t\int\limits_{\tau}^t(a+bs)^{-\beta}\,\,ds\ d\tau \le&{1\over (\beta -2)(\beta -1) b^2 a^{\beta-2}},
{\rm\ for\ }t\ge 0.&(4.9)\crcr
}
$$

For $a\ge 1,\ b>0,\ \beta>2,$
$$
\leqalignno{
\int\limits_{-\infty}^t(a+b|s|)^{-\beta}(1+|s|)\,\,ds\le&{b+1\over (\beta -2) b^2 (a-bt)^{\beta-2}},
{\rm\ for\ }t\le 0,&(4.10)\cr
\int\limits_{-\infty}^t (a+b|s|)^{-\beta}(1+|s|)\,\,ds\le&2{b+1\over (\beta -2)b^2 a^{\beta-2}},{\rm\ for\ }t\ge 0.&(4.11)\crcr
}
$$

For $a\ge 1,\ b>0,\ \beta>3,$
$$
{\int\limits_{0}^t\int\limits_{\tau}^t(a+bs)^{-\beta}(1+s)\,\,ds\ d\tau \le{b+1\over (\beta -3)(\beta -2) b^3 a^{\beta-3}},
{\rm\ for\ }t\ge 0.}\leqno(4.12)
$$

For the proof of (4.6)-(4.12), see [No].

\vskip 2mm
{\it 4.4 About $z_1(c,d,\beta_1,\alpha,|x_-|,r).$}

Let $c,\ d,\ \beta_1,\ \alpha,|x_-|,\ 0< r\le 1,\ r<c/\sqrt{2},$ be fixed. We consider the one-dimensional infinitely smooth 
function $\sigma:]\sqrt{2}r,c[
\to \R$ defined by
$$
\sigma(s)={s\over \sqrt{1-{s^2\over c^2}}}-{2^{\alpha+4}\beta_1\sqrt{d}\over \alpha({s/\sqrt{2}}-r)(|x_-|/\sqrt{2}+1)^\alpha}.
$$
$\sigma$ is an increasing function (its derivative is a positive function) and as a consequence $z_1(c,d,\beta_1,\alpha,|x_-|,r)$
is well defined in Introduction and the observation (I) of Section 3 holds.
\vskip 4mm
{\it 4.5 About $M_{T,r},\ 0< r\le 1,\ r<c/\sqrt{2}.$}

\vskip 4mm

{\bf Lemma 4.3.}
{\it Let} $f,\ f_1,\ f_2\in M_{T,r},\ v_-\in B_c\b\{0\},\ v_-x_-=0,\ |v_-|>\sqrt{2}r,$ {\it then}
$$
\ep f_1+(1-\ep) f_2 \in M_{T,r},{\it\ for\ }0\le \ep\le 1, \leqno(4.13)
$$
$$
2(1+|x_-+v_-s+f(s)|)\ge (1+|x_-|/\sqrt{2}+(|v_-|/\sqrt{2}-r)|s|),{\it \ for\ }s\le T,\leqno(4.14)
$$
$$
\left|\int_{-\infty}^t F(v_-s+x_-+f(s))ds\right|\le {\beta_1\sqrt{d}2^{\alpha+2}\over \alpha (|v_-|/\sqrt{2}-r)(|x_-|/\sqrt{2}+1)^\alpha},
{\it \ for\ }t\in ]-\infty, +\infty],\leqno(4.15)
$$
$$
\left(1+{1\over c^2}{\left|\gamma(v_-)+\ep_1 \int_{-\infty}^tF(v_-s+x_-+f_1(s))ds+\ep_2 \int_{w}^uF(v_-s+x_-+f_2(s))ds\right|^2}\right)
^{-\beta}
$$
\vskip -6mm
$$
\le
 (1+{|v_-|^2\over 4(c^2-|v_-|^2)})^{-\beta},
\leqno(4.16)
$$
{\it for} $u,t\in ]-\infty, T],\ w\in[-\infty,u],\ \beta> 0,\ -1\le\ep_1,\ep_2\le 1, f_1,f_2\in M_{T,r}$ {\it and if} 
$|v_-|\ge z_1(c,d,\beta_1,\alpha,|x_-|,r),$ $|v_-|<c,$ {\it where} $\gamma$ {\it is defined by}
$$
\gamma(v)={v\over \sqrt{1-|v|^2/c^2}},
$$
{\it for} $v\in B_c.$
\vskip 4mm
{\it Proof of Lemma} 4.3.
For the proof of (4.14) see [No]. Inequality (4.1) with (4.14) and (4.7) proves (4.15). (4.13) follows from the definition of $M_{T,r}.$ 
Inequality (4.15) 
gives in particular for $u,t\in ]-\infty, T],\ w\in[-\infty,u],\ \beta> 0,\ -1\le\ep_1,\ep_2\le 1, f_1,f_2\in M_{T,r}$
$$
\eqalign{
|\gamma(v_-)+\ep_1 \int_{-\infty}^tF(v_-s+x_-&+f_1(s))ds+\ep_2 \int_{w}^uF(v_-s+x_-+f_2(s))ds|\cr
&\ge|\gamma(v_-)|-
{\beta_1\sqrt{d}2^{\alpha+3}\over \alpha (|v_-|/\sqrt{2}-r)(|x_-|/\sqrt{2}+1)^\alpha}\cr
&={|v_-|\over\sqrt{1-|v_-|^2/c^2}}-{\beta_1\sqrt{d}2^{\alpha+3}\over \alpha (|v_-|/\sqrt{2}-r)(|x_-|/\sqrt{2}+1)^\alpha}\cr
&\ge{c|v_-|\over 2\sqrt{c^2-|v_-|^2}},{\rm \ if\ } |v_-|\ge z_1(c,d,\beta_1,\alpha,|x_-|,r),\ |v_-|<c,
}
$$
which implies (4.16).
  
\vskip 1cm

{\centerline {\bf 5. Proofs of Lemmas 2.1, 2.2, 2.3}}
\vskip 4mm
{\it Proof of Lemma} 2.1.
The property 
$$
A_{v_-,x_-}(f)\in C^1(]-\infty,T],\R^d) {\rm \ for\ }f\in M_{T,r}\ (0< r\le 1,\ r<|v_-|/\sqrt{2}) \leqno(5.1)
$$
follows from (1.2), (2.1) (applied on ``$x$"$=\gamma(v_-)+\int_{-\infty}^\tau F(v_-s+x_-+f(s))ds$ and ``$y$"$=\gamma(v_-)$) and the definition of $A_{v_-,x_-}(f).$

Now we always suppose that $0< r\le 1,$  $r<c/\sqrt{2},$ $|v_-|
\ge z_1(c,d,\beta_1,\alpha,|x_-|,r),$ $|v_-|<c,$  $v_-x_-=0.$
Consider
$$
\eqalign{
A_{v_-,x_-}(f)(t)=&\int\limits_{-\infty}^t\left[g(\gamma(v_-)+\int\limits_{-\infty}^\tau F(v_-s+x_-+f(s))ds)-v_-\right]d\tau\cr
{d\over dt}A_{v_-,x_-}(f)(t)=&g(\gamma(v_-)+\int\limits_{-\infty}^t F(v_-s+x_-+f(s))ds)-v_-.\crcr
}
{\rm \ for\ } f\in M_{T,r}.
\leqno(5.2)
$$
First we shall prove some estimates about ${d\over dt}A_{v_-,x_-}(f)$.

Note that $g(\gamma(v_-))=v_-.$
From (5.2), (4.4) (applied on ``$x$"$=\gamma(v_-)+\int_{-\infty}^t F(v_-s+x_-+f(s))ds$ and ``$y$"$=\gamma(v_-)$), (4.1), (4.14) and (4.16) 
it follows that
$$
|{d\over dt}A_{v_-,x_-}(f)(t)|\le {d \beta_12^{\alpha+1}\over \sqrt{1+{|v_-|^2\over4(c^2-|v_-|^2)}}}\int_{-\infty}^t(1+|x_-|/\sqrt{2}+
(|v_-|/\sqrt{2}-r)|s|))^{-(\alpha+1)}ds
.\leqno (5.3)
$$

Our next purpose is to prove estimates (5.5) and (5.9) given below.

From (5.3) and (4.8) and (4.6) it follows that
$$
|A_{v_-,x_-}(f)(t)|\le
{d\beta_12^{\alpha+1}\over \sqrt{1+{|v_-|^2\over 4(c^2-|v_-|^2)}}\alpha (\alpha-1)
({|v_-|\over\sqrt{2}}-r)^2(1+{|x_-|\over \sqrt{2}}-({|v_-|\over \sqrt{2}}-r)t)^{\alpha-1}},\leqno(5.4{\rm a})
$$
$$
|t{d\over dt}A_{v_-,x_-}(f)(t)|\le
{d\beta_12^{\alpha+1}\over \sqrt{1+{|v_-|^2\over 4(c^2-|v_-|^2)}}
\alpha({|v_-|\over \sqrt{2}}-r)^2(1+{|x_-|\over\sqrt{2}}-({|v_-|\over \sqrt{2}}-r)t)^{\alpha-1}},\leqno(5.4{\rm b})
$$
for $t\le T,\ t\le0$.
From (5.4) it follows that

$$
\leqalignno{
|A_{v_-,x_-}(f)(t)-t{d\over dt}&A_{v_-,x_-}(f)(t)|\cr
&\le
{d\beta_12^{\alpha+1}\over \sqrt{1+{|v_-|^2\over 4(c^2-|v_-|^2)}} (\alpha-1)
(|v_-|/\sqrt{2}-r)^2(1+|x_-|/\sqrt{2}-(|v_-|/\sqrt{2}-r)t)^{\alpha-1}},
&(5.5)
}
$$
for $t\le T,\ t\le0$.

For $t\le T,\ t\ge 0$, note that 
$$
\eqalign{
A_{v_-,x_-}(f)(t)-&t{d\over dt}A_{v_-,x_-}(f)(t)\cr
&=A_{v_-,x_-}(f)(0)+
\int\limits_0^t\left[g(\gamma(v_-)+\int\limits_{-\infty}^\tau F(v_-s+x_-+f(s))ds)\right.\cr
&-\left.g(\gamma(v_-)+\int\limits_{-\infty}^t F(v_-s+x_-+f(s))ds)\right]d\tau.\crcr
}
\leqno(5.6)
$$
For $A_{v_-,x_-}(f)(0)$ we use the estimate (5.4a), i.e.
$$
|A_{v_-,x_-}(f)(0)|\le
{d\beta_12^{\alpha+1}\over \sqrt{1+(|v_-|^2/(4(c^2-|v_-|^2)))}\alpha (\alpha-1)
(|v_-|/\sqrt{2}-r)^2(1+|x_-|/\sqrt{2})^{\alpha-1}}.
\leqno(5.7)
$$
We estimate the second term on the right-hand side of (5.6) in the following way: from (4.4), (4.16), (4.1), (4.14) and (4.9),  
it follows that
$$
\left|\int\limits_0^t\left[g(\gamma(v_-)+\int\limits_{-\infty}^\tau F(v_-s+x_-+f(s))ds)
-g(\gamma(v_-)+\int\limits_{-\infty}^t F(v_-s+x_-+f(s))ds)\right]d\tau\right|
$$
\vskip -2mm
$$
\leqalignno{\le &{\sqrt{d}\over \sqrt{1+{|v_-|^2\over4(c^2-|v_-|^2)}}}
\int^t_0\left|\int_{t}^{\tau}F(v_-s+x_-+f(s))ds\right|d\tau\cr
\le &{d\beta_12^{\alpha+1}\over \sqrt{1+(|v_-|^2/(4(c^2-|v_-|^2)))}\alpha (\alpha-1)
(|v_-|/\sqrt{2}-r)^2(1+|x_-|/\sqrt{2})^{\alpha-1}},
&(5.8)
}
$$
for $0\le t\le T.$
From (5.6), (5.7) and (5.8) it follows that
$$
\leqalignno{|A_{v_-,x_-}(f)(t)-t&{d\over dt}A_{v_-,x_-}(f)(t)|
\cr
&\le 
{d\beta_12^{\alpha+2}\over \sqrt{1+{|v_-|^2\over(4(c^2-|v_-|^2)}}\alpha(\alpha-1)
(|v_-|/\sqrt{2}-r)^2(1+|x_-|/\sqrt{2})^{\alpha-1}}
&(5.9)
}
$$
for $0\le t\le T.$
Using (5.3) and (4.6) and using (5.5) we obtain (2.7a).
Using (5.3) and (4.7) and using (5.9) we obtain (2.7b).

Our next purpose is to prove estimate (5.14) given below.
Consider ${d\over dt}(A_{v_-,x_-}(f_2)(t)-A_{v_-,x_-}(f_1)(t))$ for 
$f_1,\ f_2 \in M_{T,r}$ $(0< r\le 1,$ $r<c/\sqrt{2},$ $|v_-|<c,$ $v_-x_-=0,$ $|v_-|\ge z_1(c,d,\beta_1,\alpha,|x_-|,r)).$
First
$$
\leqalignno{
{d\over dt}A_{v_-,x_-}(f_2)(t)&-{d\over dt}A_{v_-,x_-}(f_1)(t)
\cr
&=g(\gamma(v_-)+\int\limits_{-\infty}^t F(v_-s+x_-+f_2(s))ds)
-g(\gamma(v_-)+\int\limits_{-\infty}^t F(v_-s+x_-+f_1(s))ds)
&(5.10)
}
$$
for $t\le T.$
From (5.10), (4.4) and (4.16)  it follows that
$$
\leqalignno{
|{d\over dt}(A_{v_-,x_-}(f_2)(t)&-{d\over dt}A_{v_-,x_-}(f_1)(t))|\cr
&\le {\sqrt{d} \over \sqrt{1+(|v_-|^2/(4(c^2-|v_-|^2)))}}\int\limits_{-\infty}^t |F(v_-s+x_-+f_2(s))-F(v_-s+x_-+f_1(s))|ds,
&(5.11)
}
$$
for $t\le T.$
From (4.13), (4.14) and (4.2), it follows that 
$$
\leqalignno{
|F(v_-s+x_-+f_2(s))&-F(v_-s+x_-+f_1(s))|\cr
&\le d\beta_2 2^{\alpha+2}(1+|x_-|/\sqrt{2}+(|v_-|/\sqrt{2}-r)|s|)^{-(\alpha+2)}|f_2(s)-f_1(s)|,
 {\rm\ for\ }s\le T.
&(5.12)
}
$$  
Moreover
$$
|f_2(s)-f_1(s)|\le (1+|s|)\|f_2-f_1\|_T,{\rm\ for\ }s\le T.
\leqno(5.13)
$$
Thus, from (5.11), (5.12) and (5.13) it follows that
$$
\leqalignno{
|{d\over dt}A_{v_-,x_-}(f_2)(t)&-{d\over dt}A_{v_-,x_-}(f_1)(t)|\cr
&\le {d\sqrt{d}\beta_22^{\alpha+2}\|f_2-f_1\|_T\over
\sqrt{1+(|v_-|^2/(4(c^2-|v_-|^2)))}}\int_{-\infty}^t(1+|x_-|/\sqrt{2}+(|v_-|/\sqrt{2}-r)|s|)^{-(\alpha+2)}(1+|s|)ds.
&(5.14)
}
$$

Our next purpose is to prove estimates (5.17) and (5.31) given below.
From (5.14) and (4.10) and (4.6) it follows that
$$
\leqalignno{
|A_{v_-,x_-}&(f_2)(t)-A_{v_-,x_-}(f_1)(t)|\cr
&\le
{d\sqrt{d}\beta_22^{\alpha+2}(|v_-|/\sqrt{2}+1-r)\|f_2-f_1\|_T\over
\sqrt{1+{|v_-|^2\over4(c^2-|v_-|^2)}}\alpha (\alpha-1)(|v_-|/\sqrt{2}-r)^3(1+|x_-|/\sqrt{2}-(|v_-|/\sqrt{2}-r)t)^{\alpha-1}}
&(5.15)
}
$$
for $t\le T,\ t\le 0.$ 
From (5.14) and (4.10) it also follows that
$$
\leqalignno{
|t||{d\over dt}&A_{v_-,x_-}(f_2)(t)-{d\over dt}A_{v_-,x_-}(f_1)(t)|\cr
&\le {d\sqrt{d}\beta_22^{\alpha+2}(|v_-|/\sqrt{2}+1-r)\|f_2-f_1\|_T\over
\sqrt{1+{|v_-|^2\over 4(c^2-|v_-|^2)}}\alpha (|v_-|/\sqrt{2}-r)^3(1+|x_-|/\sqrt{2}-(|v_-|/\sqrt{2}-r)t)^{\alpha-1}}
&(5.16)
}
$$
for $t\le T,\ t\le 0.$ 
Hence from (5.15) and (5.16) it follows that
$$
\leqalignno{
|A_{v_-,x_-}&(f_2)(t)-A_{v_-,x_-}(f_1)(t)-t{d\over dt}(A_{v_-,x_-}(f_2)(t)-A_{v_-,x_-}(f_1)(t))|\cr
&\le 
{d\sqrt{d}\beta_22^{\alpha+2}(|v_-|/\sqrt{2}+1-r)\|f_2-f_1\|_T\over
\sqrt{1+{|v_-|^2\over4(c^2-|v_-|^2)}}(\alpha-1) (|v_-|/\sqrt{2}-r)^3(1+|x_-|/\sqrt{2}-(|v_-|/\sqrt{2}-r)t)^{\alpha-1}}
&(5.17)
}
$$
for $t\le T,t\le 0$.

For $0\le t\le T$, using (5.6) we obtain 
$$
|A_{v_-,x_-}(f_2)(t)-A_{v_-,x_-}(f_1)(t)-t({d\over dt}A_{v_-,x_-}(f_2)(t)-{d\over dt}A_{v_-,x_-}(f_1)(t))|
$$
\vskip -4mm
$$
\eqalign{
&\le|A_{v_-,x_-}(f_2)(0)-A_{v_-,x_-}(f_1)(0)|\cr
&+\left|\int\limits_0^t\left[g(\gamma(v_-)+\int\limits_{-\infty}^\tau F(v_-s+x_-+f_2(s))ds)-g(\gamma(v_-)+\int\limits_{-\infty}^t
F(v_-s+x_-+f_2(s))ds)\right.\right.\cr
&\left.\left.-g(\gamma(v_-)+\int\limits_{-\infty}^{\tau} F(v_-s+x_-+f_1(s))ds)+g(\gamma(v_-)+\int\limits_{-\infty}^t
 F(v_-s+x_-+f_1(s))ds)
\right]d\tau\right|.\crcr
}
\leqno(5.18)
$$
From (5.15) it follows that
$$
|A_{v_-,x_-}(f_2)(0)-A_{v_-,x_-}(f_1)(0)|\le
{d\sqrt{d}\beta_22^{\alpha+2}(|v_-|/\sqrt{2}+1-r)\|f_2-f_1\|_T\over
\sqrt{1+{|v_-|^2\over4(c^2-|v_-|^2)}}\alpha (\alpha-1)( {|v_-|\over\sqrt{2}}-r)^3(1+{|x_-|\over\sqrt{2}})^{\alpha-1}}
.\leqno(5.19)
$$
In order to estimate the second term of the right-hand side of (5.18), we will estimate 
$$ 
\int\limits_0^t\left[g_j(\gamma(v_-)+\int\limits_{-\infty}^\tau F(v_-s+x_-+f_2(s))ds)-g_j(\gamma(v_-)+\int\limits_{-\infty}^t
F(v_-s+x_-+f_2(s))ds)\right.
$$
\vskip -4mm
$$
\left.-g_j(\gamma(v_-)+\int\limits_{-\infty}^{\tau} F(v_-s+x_-+f_1(s))ds)+g_j(\gamma(v_-)+\int\limits_{-\infty}^t
 F(v_-s+x_-+f_1(s))ds)
\right]d\tau
\leqno(5.20)
$$
for $1\le j\le d$ and $0\le t\le T$.

Let $1\le j\le d$ and $0\le t\le T,\ 0\le \tau\le t$. 
Note that
$$
g_j(\gamma(v_-)+\int\limits_{-\infty}^{\tau} F(v_-s+x_-+f_2(s))ds)-g_j(\gamma(v_-)+\int\limits_{-\infty}^t
 F(v_-s+x_-+f_2(s))ds)
$$
$$
-\left(g_j(\gamma(v_-)+\int\limits_{-\infty}^{\tau} F(v_-s+x_-+f_1(s))ds)-g_j(\gamma(v_-)+\int\limits_{-\infty}^t
 F(v_-s+x_-+f_1(s))ds)
\right)
$$
\vskip -4mm
$$
=\Delta^1_{j,t}(\tau)+\Delta^2_{j,t}(\tau)
\leqno(5.21)
$$
where
$$
\Delta^1_{j,t}(\tau)=\int_t^{\tau}(F(v_-s+x_-+f_2(s))-F(v_-s+x_-+f_1(s)))ds
\leqno(5.22{\rm a})
$$
\vskip -4mm
$$
\star\int_0^1\nabla g_j\left(\gamma(v_-)+\int\limits_{-\infty}^{t} F(v_-s+x_-+f_2(s))ds
+\ep \int\limits_{t}^{\tau} F(v_-s+x_-+f_2(s))ds\right)d\ep,
$$
\vskip -4mm
$$
\Delta^2_{j,t}(\tau)=\int_t^{\tau}F(v_-s+x_-+f_1(s))ds
\leqno(5.22{\rm b})
$$
\vskip -4mm
$$
\star\int_0^1\left[ \nabla g_j(\gamma(v_-)+\int\limits_{-\infty}^{t} F(v_-s+x_-+f_2(s))ds
+\ep \int\limits_{t}^{\tau} F(v_-s+x_-+f_2(s))ds)\right.
$$
\vskip -4mm
$$
\left.-\nabla g_j(\gamma(v_-)+\int\limits_{-\infty}^{t} F(v_-s+x_-+f_1(s))ds
+\ep \int\limits_{t}^{\tau} F(v_-s+x_-+f_1(s))ds)\right]d\ep.
$$

Using (5.22a), (4.3), (4.2), (4.16), (4.14) and (5.13), we obtain 
$$
|\Delta^1_{j,t}(\tau)|\le {d\beta_22^{\alpha+2}\over
\sqrt{1+{|v_-|^2\over4(c^2-|v_-|^2)}}}\int_{\tau}^t(1+|x_-|/\sqrt{2}+(|v_-|/\sqrt{2}-r)s)^{-(\alpha+2)}(1+s)ds\|f_2-f_1\|_T.
\leqno(5.23)
$$
Thus from (4.12) it follows that
$$
\int_0^t|\Delta^1_{j,t}(\tau)|d\tau\le {d\beta_22^{\alpha+2}(|v_-|/\sqrt{2}+1-r)\over
\sqrt{1+{|v_-|^2\over4(c^2-|v_-|^2)}}\alpha(\alpha-1)(|v_-|/\sqrt{2}-r)^3(1+|x_-|/\sqrt{2})^{\alpha-1}}\|f_2-f_1\|_T.
\leqno(5.24)
$$
Using (5.22b), (4.5) and (4.16), we obtain
$$
\eqalign{
|\Delta^2_{j,t}(\tau)|\le&\int_{\tau}^t|F(v_-s+x_-+f_1(s))|ds\left[{3\sqrt{d}\over c(1+(|v_-|^2/(4(c^2-|v_-|^2))))}\right.\cr
&\times\int_0^1\left|\int_{-\infty}^t(F(v_-s+x_-+f_2(s))-F(v_-s+x_-+f_1(s)))ds\right.\cr
&\left.\left.+\ep \int_t^{\tau}(F(v_-s+x_-+f_2(s))-F(v_-s+x_-+f_1(s)))ds\right|d\ep\right].\crcr
}
\leqno(5.25)
$$
We shall use
$$
\leqalignno{
\left|\int_{-\infty}^t(F(v_-s+\right.&\left.x_-+f_2(s))-F(v_-s+x_-+f_1(s)))ds\right.
\cr
\left.+\ep \int_t^{\tau}(F(v_-s+\right.&\left.x_-+f_2(s))-F(v_-s+x_-+f_1(s)))ds\right|
\cr
&\le 2\int_{-\infty}^t|(F(v_-s+x_-+f_2(s))-F(v_-s+x_-+f_1(s)))|ds,
&(5.26)
}
$$
for all $0\le \ep\le 1$ (we remind that $\tau\le t$).

From (5.25) and (5.26) it follows that
$$
\leqalignno{
|\Delta^2_{j,t}(\tau)|\le&
\int_{\tau}^t|F(v_-s+x_-+f_1(s))|ds\cr
&\times{6\sqrt{d}\over c(1+{|v_-|^2\over 4(c^2-|v_-|^2)})}
\int_{-\infty}^t|(F(v_-s+x_-+f_2(s))-F(v_-s+x_-+f_1(s)))|ds.
&(5.27)
}
$$
Using (4.2), (4.14), (5.13) and (4.11) we obtain
$$
\leqalignno{
\int_{-\infty}^t|F(v_-s+&x_-+f_2(s))-F(v_-s+x_-+f_1(s))|ds\cr
& \le {d\beta_22^{\alpha+3}(|v_-|/\sqrt{2}+1-r)
\over \alpha(|v_-|/\sqrt{2}-r)^2(1+|x_-|/\sqrt{2})^{\alpha}}\|f_2-f_1\|_T.
&(5.28)
}
$$
Using (4.1), (4.14) and (4.9) we obtain
$$
\int_0^t\int_{\tau}^t|F(v_-s+x_-+f_1(s))|dsd\tau\le {\sqrt{d}\beta_12^{\alpha+1}\over
\alpha(\alpha-1)(|v_-|/\sqrt{2}-r)^2(1+|x_-|/\sqrt{2})^{\alpha-1}}.
\leqno(5.29)
$$
From (5.27), (5.28) and (5.29) it follows that
$$
\leqalignno{
\int_0^t|\Delta^2_{j,t}(\tau)|d\tau\le&{3\over c(1+(|v_-|^2/(4(c^2-|v_-|^2))))}\cr
&\times{d^2\beta_1\beta_22^{2\alpha+5}(|v_-|/\sqrt{2}+1-r)\over \alpha^2(\alpha-1)(|v_-|/\sqrt{2}-r)^4(1+|x_-|/\sqrt{2})^{2\alpha-1}}
\|f_2-f_1\|_T.&(5.30)\cr
}
$$
From (5.18), (5.19), (5.21), (5.24) and (5.30), it follows that
$$
|A_{v_-,x_-}(f_2)(t)-A_{v_-,x_-}(f_1)(t)-t({d\over dt}A_{v_-,x_-}(f_2)(t)-{d\over dt}A_{v_-,x_-}(f_1)(t))|
$$
\vskip -4mm
$$
\leqalignno{
\le& {d\sqrt{d}\beta_22^{\alpha+2}(|v_-|/\sqrt{2}+1-r)\over\sqrt{1+(|v_-|^2/(4(c^2-|v_-|^2)))}
\alpha (\alpha-1)(|v_-|/\sqrt{2}-r)^3(1+|x_-|/\sqrt{2})^{\alpha-1}}\cr
&\times \left[2+{3\over c\sqrt{(1+(|v_-|^2/(4(c^2-|v_-|^2))))}}
{d\beta_12^{\alpha+3}\over \alpha(|v_-|/\sqrt{2}-r)(1+|x_-|/\sqrt{2})^{\alpha}}\right]&(5.31)\cr
&\times\|f_1-f_2\|_T.\crcr
}
$$
Using (5.14) and (4.10) and (5.17) we obtain (2.8a).
Using (5.14) and (4.11) and (5.31) we obtain (2.8b). 

Lemma 2.1 is proved. 

{\it Proof of Lemma 2.2.} The estimates (2.10) and (2.11) follow immediately from (5.3) and (4.6) and (5.4a).
From (4.1), (4.14) and (4.7) it follows that 
$$
\int\limits_{-\infty}^{+\infty} F(v_-s+x_-+f(s))ds 
$$
converges absolutely for any $f\in M_{T,r}.$
Moreover, using (4.4), (4.16) and then (4.1), (4.14) and (4.8) we obtain for $u>0$
$$
\int_u^{+\infty} \left|g(\gamma(v_-)+\int\limits_{-\infty}^\tau F(v_-s+x_-+f(s))ds)-
		  g(\gamma(v_-)+\int\limits_{-\infty}^{+\infty} F(v_-s+x_-+f(s))ds)\right|d\tau
$$
\vskip -6mm
$$
\leqalignno{
\le& {d\beta_12^{\alpha+1} \over \sqrt{1+{|v_-|^2\over4(c^2-|v_-|^2)}}} \int_u^{+\infty }\!\int_{\tau}^{+\infty}(1+|x_-|/\sqrt{2}+
(|v_-|/\sqrt{2}-r)s)^{-(\alpha+1)}dsd\tau\cr
\le&{d\beta_12^{\alpha+1} (1+|x_-|/\sqrt{2}+(|v_-|/\sqrt{2}-r)u)^{-(\alpha-1)}\over
\sqrt{1+{|v_-|^2\over4(c^2-|v_-|^2)}}\alpha(\alpha-1)(|v_-|/\sqrt{2}-r)^2}.&(5.32)\cr
}
$$
As a consequence we can write
$$
\leqalignno{
A_{v_-,x_-}(f)(t)=&t\left[g(\gamma(v_-)+\int\limits_{-\infty}^{+\infty} F(v_-s+x_-+f(s))ds)-v_-\right]\cr
		  &+\int\limits_{-\infty}^0\left[g(\gamma(v_-)+\int\limits_{-\infty}^\tau F(v_-s+x_-+f(s))ds)-v_-\right]d\tau\cr    
                  &+\int\limits_0^{+\infty}\left[g(\gamma(v_-)+\int\limits_{-\infty}^\tau\!\!\! F(v_-s+x_-+f(s))ds)-
		  g(\gamma(v_-)+\int\limits_{-\infty}^{+\infty}\!\!\! F(v_-s+x_-+f(s))ds)\right]d\tau\cr
                  &-\int\limits_t^{+\infty}\left[g(\gamma(v_-)+\int\limits_{-\infty}^\tau\!\!\! F(v_-s+x_-+f(s))ds)-
		  g(\gamma(v_-)+\int\limits_{-\infty}^{+\infty}\!\!\!F(v_-s+x_-+f(s))ds)\right]d\tau&(5.33)\crcr
}
$$
and (2.12) and (2.13) follow, where
$$
\leqalignno{
H_{v_-,x_-}(f)(t)=&\int\limits_t^{+\infty}\left[g(\gamma(v_-)+\int\limits_{-\infty}^{+\infty} F(v_-s+x_-+f(s))ds)\right.\cr
&\left.-g(\gamma(v_-)+\int\limits_{-\infty}^{\tau} F(v_-s+x_-+f(s))ds)\right]d\tau.&(5.34)
}
$$
The formulas (5.34) and (5.32) prove (2.16).
Using (5.34), (4.4), (4.16), (4.1), (4.14) and (4.6), we obtain (2.15).

Using (2.13a), (4.4), (4.16), (4.1), (4.14) and (4.7) we obtain (2.14a).

We write
$$
\leqalignno{
l_{v_-,x_-}(f)=& A_{v_-,x_-}(f)(0)
	       +\int\limits_0^{+\infty}\left[g(\gamma(v_-)+\int\limits_{-\infty}^\tau F(v_-s+x_-+f(s))ds)\right.\cr
	      &\left.-g(\gamma(v_-)+\int\limits_{-\infty}^{+\infty} F(v_-s+x_-+f(s))ds)\right]d\tau.&(5.35)\crcr              
}
$$
Using (5.35), (5.7) and (5.32), we obtain (2.14b).

Thus Lemma 2.2 is proved.

\vskip 2mm

{\it Proof of Lemma 2.3.} 
Using (2.3) and (2.7b) we obtain
$$
\|y_--0\|_T=\|y_-\|_T\le \rho(c,d,\beta_1,\alpha,|v_-|,|x_-|,r),\ T=+\infty.
$$
Using (2.13a), (5.10) with (5.14) and (4.11) ($T=+\infty$ and $t\to +\infty$), we obtain (2.17a).

From (5.35) it follows that
$$
|l_{v_-,x_-}(y_-)-l_{v_-,x_-}(0)|\le|A_{v_-,x_-}(y_-)(0)-A_{v_-,x_-}(0)(0)|\hfill \leqno(5.36)
$$

\vskip -8mm

$$
\leqalignno{
   &+\left|\lim_{t\to +\infty}\left\lbrace\int_0^t\left[g(\gamma(v_-)+\int_{-\infty}^\tau F(v_-s+x_-+y_-(s))ds)-
   g(\gamma(v_-)+\int_{-\infty}^t F(v_-s+x_-+y_-(s))ds)\right.\right.\right.\cr
   &-g(\gamma(v_-)+\int_{-\infty}^\tau F(v_-s+x_-)ds)+g(\gamma(v_-)+\int_{-\infty}^t F(v_-s+x_-)ds)\cr
   &-\left(g(\gamma(v_-)+\int_{-\infty}^{+\infty} F(v_-s+x_-+y_-(s))ds)-g(\gamma(v_-)+\int_{-\infty}^t
   F(v_-s+x_-+y_-(s))ds)\right)\cr
   &+\left.\left.\left.\left(g(\gamma(v_-)+\int_{-\infty}^{+\infty} F(v_-s+x_-)ds)-g(\gamma(v_-)+\int_{-\infty}^t F(v_-s+x_-)ds)\right)
   \right]d\tau\right\rbrace\right|.\crcr    
}
$$
Using (4.4), (4.16), (4.1), (4.14) and (4.6) we obtain
$$
\leqalignno{
t\left|g(\gamma(v_-)+\int_{-\infty}^{+\infty} F(v_-s+x_-+f(s))ds)\right.&\left.-g(\gamma(v_-)+\int_{-\infty}^t F(v_-s+x_-+f(s))ds)\right|\cr
&\to 0\ {\rm as\ }t\to +\infty &(5.37)
}
$$
for $f\in M_{T,r}.$

From (5.36) and (5.37) it follows that
$$
|l_{v_-,x_-}(y_-)-l_{v_-,x_-}(0)|\le|A_{v_-,x_-}(y_-)(0)-A_{v_-,x_-}(0)(0)|\leqno(5.38)
$$

\vskip -8mm

$$
\leqalignno{
   &+\left|\lim_{t\to +\infty}\left\lbrace\int_0^t\left[g(\gamma(v_-)+\int_{-\infty}^\tau F(v_-s+x_-+y_-(s))ds)-
   g(\gamma(v_-)+\int_{-\infty}^t F(v_-s+x_-+y_-(s))ds)\right.\right.\right.\cr
   &\left.\left.\left.-g(\gamma(v_-)+\int_{-\infty}^\tau F(v_-s+x_-)ds)+g(\gamma(v_-)+\int_{-\infty}^t F(v_-s+x_-)ds)
   \right]d\tau\right\rbrace\right|.\crcr
}
$$
Using (5.19), (5.20), (5.21), (5.24), (5.30), $\|y_-\|_T$ $\le$ $\rho(c,d,\beta_1,\alpha,|v_-|,|x_-|,r),$  $T=+\infty,$  
and (5.38) we obtain (2.17c).

We shall prove (2.17b). First
$$
v_-+k_{v_-,x_-}(y_-)=g(\gamma(v_-)+\int_{-\infty}^{+\infty} F(v_-s+x_-+y_-(s))ds).
\leqno(5.39)
$$
Using the integral of motion $E$, we have $|v_-|=|v_-+k_{v_-,x_-}(y_-)|$ and applying $\gamma$ to (5.39) we obtain
$$
{k_{v_-,x_-}(y_-)\over\sqrt{1-|v_-|^2/c^2}}=\int_{-\infty}^{+\infty} F(v_-s+x_-+y_-(s))ds.
\leqno(5.40)
$$
From (5.40), (5.12), (5.13) and (4.11) and $\|y_-\|_T\le \rho(c,d,\beta_1,\alpha,|v_-|,|x_-|,r),\ T=+\infty,$ we obtain (2.17b)

Lemma 2.3 is proved.

\vskip 1cm

{\centerline {\bf 6. Proofs of Theorems 3.2 and Proposition 1.1}} 

\vskip 2mm

Let $(\theta, x)\in T\S^{d-1},\ \alpha,\ d,\ c,\ \beta_1,\ \beta_2$ be fixed.

We shall use 
$$
\left|{1\over s}\int_{-\infty}^uF(x+\tau \theta)d\tau\right|\le {\beta_1\sqrt{d}\over \alpha(s/\sqrt{2})(1+|x|/\sqrt{2})^\alpha}
\leqno(6.1{\rm a})
$$
for $s\in]0,c[$ and $u\in]-\infty,0]$; replacing $\theta$ by $-\theta$ in (6.1a), we obtain
$$
\left|{1\over s}\int^{+\infty}_uF(x+\tau \theta)d\tau\right|\le {\beta_1\sqrt{d}\over \alpha(s/\sqrt{2})(1+|x|/\sqrt{2})^\alpha}
\leqno(6.1{\rm b})
$$
for $s\in]0,c[$ and $u\in[0,+\infty[$.

We prove (6.1a).
As $\theta x=0$, the following formula is valid:
$$
|x+w\theta|\ge |x|/\sqrt{2}+|w|/\sqrt{2},\leqno(6.2)
$$ 
for any $w\in \R.$
Then estimate (6.1a) follows from (4.1), (6.2) and (4.6).

Before proving Theorem 3.2, we need introduce three Lemmas and prove them.  
\vskip 2mm

{\bf Lemma 6.1.} {\it There exists integrable} $\tilde{g}_{c,d,\beta_0,\beta_1,\alpha,|x|}:]-\infty,0]\to [0,+\infty[$ {\it such that
$$
\left|
  \left(1+\delta_1(c,\theta,x,s,u)\right)^{-{1 \over2}} 
-1-{V(x+u\theta)\sqrt{1- {s^2\over c^2}}\over c^2}
\right|
\le 
\tilde{g}_{c,d,\beta_0,\beta_1,\alpha,|x|}(u)(1-s^2/c^2),
\leqno(6.3)
$$
for $u\in ]-\infty,0]$ and $s<c,\ s\ge z_2(c,d,\beta_1,\alpha,|x|),$ and where
$$
\delta_1(c,\theta,x,s,u)
={-2V(x+u\theta)\sqrt{1-{s^2\over c^2}}+({1\over s^2}-{1\over c^2})\left|\int^u_{-\infty}F(x+\tau\theta)d\tau\right|^2\over c^2}\ge
-{3\over4}
,\leqno(6.4)
$$
for $u\in ]-\infty,0]$ and $s<c,\ s\ge z_2(c,d,\beta_1,\alpha,|x|).$
}
\vskip 4mm
{\it Proof of Lemma} 6.1. 

Let $s\in]0,c[, s\ge z_2(c,d,\beta_1,\alpha,|x|) $ and $u\in]-\infty,0]$.

From (6.1a) and the definition of $z_2(c,d,\beta_1,\alpha,|x|)$ (see (1.7d)) it follows that
$$
\left|{s\theta\over \sqrt{1-s^2/c^2}}+{1\over s}\int_{-\infty}^uF(x+\tau \theta)d\tau\right|
\ge 
{s\over 2\sqrt{1-s^2/c^2}}.
\leqno(6.5)
$$ 

Expanding the square of the norm we obtain: 
$$
\left|{s\theta\over \sqrt{1-s^2/c^2}}+{1\over s}\int_{-\infty}^uF(x+\tau \theta)d\tau\right|^2
={s^2\over 1-s^2/c^2}-{2V(x+u\theta)\over \sqrt{1-s^2/c^2}}+\left|{1\over s}\int_{-\infty}^uF(x+\tau \theta)d\tau\right|^2.
\leqno(6.6)
$$
Using (6.5) and (6.6), we obtain
$$
\leqalignno{
\delta_1(c,\theta,x,s,u)=&{1+{1\over c^2}\left|{s\theta\over \sqrt{1-s^2/c^2}}+{1\over s}\int_{-\infty}^uF(x+\tau \theta)d\tau\right|^2
\over 1+{s^2\over c^2-s^2 }}-1
\ge{1+{s^2\over 4(c^2-s^2)}\over 1+{s^2\over c^2-s^2}}-1\cr
\ge&-3/4.&(6.7)}
$$
Moreover, from the definition of $\delta_1(c,\theta,x,s,u),$ (1.2), (6.1a) and the hypothesis 
$s\ge z_2(c,d,\beta_1,\alpha,|x|),$ $s<c,$ it follows that
$$
\leqalignno{
|\delta_1(c,\theta,x,s,u)|
   \le&\sqrt{1-s^2/c^2}\left[{\beta_02(1+|x|/\sqrt{2}-u/\sqrt{2})^{-\alpha}\over c^2}\right.\cr
   &\left.+{\sqrt{1-z_2(c,d,\beta_1,\alpha,|x|)^2/c^2}\beta_1^2d 2 (1+|x|/\sqrt{2}-u/\sqrt{2})^{-2\alpha}
   \over z_2(c,d,\beta_1,\alpha,|x|)^2c^2\alpha^2}
                       \right].&(6.8)   
}
$$ 

Using Taylor expansion of the map $]-1,+\infty[\to \R$, $\delta\mapsto (1+\delta)^{-1/2}$ at $\delta=0$, we obtain that 
 
$$
\leqalignno{
\left(1+\delta_1(c,\theta,x,s,u)\right)^{-{1 \over2}} 
&-1-{V(x+u\theta)\sqrt{1- {s^2\over c^2}}\over c^2}=-{1-s^2/c^2\over2s^2c^2}\left|\int_{-\infty}^uF(x+\tau\theta)d\tau\right|^2\cr
&+{3\over4}\int_0^1(1-w)(1+w\delta_1(c,\theta,x,s,u))^{-5/2}dw\ \delta_1(c,\theta,x,s,u)^2.&(6.9)
}
$$

We estimate the first term of the right-hand side of (6.9) with the help of (6.1a).We estimate the second term of the right-hand side
of (6.9) with the help of (6.7) and (6.8). Using also the inequality $s\ge z_2(c,d,\beta_1,\alpha,|x|),$ we finally obtain
$$
\left|\left(1+\delta_1(c,\theta,x,s,u)\right)^{-{1 \over2}} 
-1-{V(x+u\theta)\sqrt{1- {s^2\over c^2}}\over c^2}\right|
\le
\tilde{g}_{c,d,\beta_0,\beta_1,\alpha,|x|}(u)
$$
where
$$
\eqalign{
\tilde{g}_{c,d,\beta_0,\beta_1,\alpha,|x|}(u)
=&
{d\beta_1^2\over c^2z_2(c,d,\beta_1,\alpha,|x|)^2\alpha^2(1+|x|/\sqrt{2}-u/\sqrt{2})^{2\alpha}}
\cr
&+4^{5/2}{3\over2c^4}\left[\beta_0(1+|x|/\sqrt{2}-u/\sqrt{2})^{-\alpha}\right.\cr
&\left.+{\sqrt{1-z_2(c,d,\beta_1,\alpha,|x|)^2/c^2}\beta_1^2d(1+|x|/\sqrt{2}-u/\sqrt{2})^{-2\alpha}
   \over z_2(c,d,\beta_1,\alpha,|x|)^2\alpha^2}\right]^2.
\crcr
}
$$
Lemma 6.1 is proved.
\vskip 4mm

{\bf Lemma 6.2.} 
{\it Let} $\beta>0,\ s\in]0,c[, s\ge z_2(c,d,\beta_1,\alpha,|x|).$ {\it Then there exists a positive real number}
 $k_{\beta,c,d,\beta_1,\alpha,|x|}$ {\it such that}
$$
\left|\left(1+{1-s^2/c^2 \over s^2c^2}\left|\int_{-\infty}^{+\infty}F(x+\tau \theta)d\tau\right|^2
       \right)^{-\beta}-1  
\right|
\le 
(1-s^2/c^2)k_{\beta,c,d,\beta_1,\alpha,|x|}.
$$

\vskip 4mm

{\it Proof of Lemma} 6.2. 
We define 
$$
\delta_2(c,\theta,x,s)={1-s^2/c^2 \over s^2c^2}\left|\int_{-\infty}^{+\infty}F(x+\tau \theta)d\tau\right|^2\ge0.
\leqno(6.10)
$$
Using (6.1) and $s\ge z_2(c,d,\beta_1,\alpha,|x|),$ we obtain
$$
\delta_2(c,\theta,x,s)\le(1-s^2/c^2){d\beta_1^28\over c^2z_2(c,d,\beta_1,\alpha,|x|)^2\alpha^2(1+|x|/\sqrt{2})^{2\alpha}}.
\leqno(6.11)
$$
Using the Taylor expansion of the map $]-1,+\infty[\to \R$, $\delta\mapsto (1+\delta)^{-\beta}$ at $\beta=0$ and using (6.10), we obtain
$$
(1+\delta_2(c,\theta,x,s))^{-\beta}-1=-\beta\delta_2(c,\theta,x,s)\int_0^1(1+w\delta_2(c,\theta,x,s))^{-(\beta+1)}dw.
\leqno(6.12)
$$
From (6.10), (6.11) and (6.12) it follows that
$$
\eqalign{
|(1+\delta_2(c,\theta,x,s))^{-\beta}-1|\le& \beta \delta_2(c,\theta,x,s)
\le (1-s^2/c^2)k_{\beta,c,d,\beta_1,\alpha,|x|},
}
$$
where 
$$
k_{\beta,c,d,\beta_1,\alpha,|x|}={\beta d\beta_1^28\over c^2z_2(c,d,\beta_1,\alpha,|x|)^2\alpha^2(1+|x|/\sqrt{2})^{2\alpha}}.
$$
Lemma 6.2 is proved.
\vskip 4mm

We always suppose that $(\theta, x)\in T\S^{d-1},\ \alpha,\ d,\ c,\ \beta_1,\ \beta_2$ are fixed. Let $s\in]0,c[, s\ge z_2(c,d,\beta_1,\alpha,|x|),\ u
\in [0,+\infty[$. we define
$$
A(c,\theta,x,s,u)= \left(1+t(c,\theta,x,s,u)\right)^{-{1\over2}}.\leqno(6.13)
$$
where
$$
t(c,\theta,x,s,u)={1+{1\over c^2}\left|{s\theta\over\sqrt{1-s^2/c^2}}+{1\over s}\int_{-\infty}^u F(x+\tau\theta)d\tau\right|^2
\over 1+{1\over c^2}\left|{s\theta\over\sqrt{1-s^2/c^2}}+{1\over s}\int_{-\infty}^{+\infty} F(x+\tau\theta)d\tau\right|^2}-1.
\leqno(6.14)
$$
Expanding square of the norms in the numerator and denominator of the fraction of the right-hand side of (6.14), we obtain that
$$
\eqalign{
t(c,\theta,x,s,u)=&{-2V(x+u\theta)\sqrt{1-s^2/c^2}+{1-s^2/c^2\over s^2}\left|\int_u^{+\infty}F(x+\tau \theta)d\tau\right|^2
\over \left(1+{(1-s^2/c^2)\over s^2c^2}\left|\int_{-\infty}^{+\infty}F(x+\tau \theta)d\tau\right|^2\right)c^2}\cr
&-{{2(1-s^2/c^2)\over s^2}\int_u^{+\infty}F(x+\tau \theta)d\tau.\int_{-\infty}^{+\infty}F(x+\tau \theta)d\tau
\over \left(1+{(1-s^2/c^2)\over s^2c^2}\left|\int_{-\infty}^{+\infty}F(x+\tau \theta)d\tau\right|^2\right)c^2
}.
}
\leqno(6.15)
$$
\vskip 4mm
{\bf Lemma 6.3.} {\it There exists} $h_{c,d,\beta_0,\beta_1,\alpha,|x|}:[0,+\infty[\to [0,+\infty[$ {\it an integrable function such that for }
$s\in]0,c[,\ s\ge z_2(c,d,\beta_1,\alpha,|x|),\ u\in [0,+\infty[,$
$$
\left|A(c,\theta,x,s,u)-1-V(x+u\theta){\sqrt{1-s^2/c^2}\over c^2}\right|\le (1-s^2/c^2)h_{c,d,\beta_0,\beta_1,\alpha,|x|}(u).
$$
\vskip 4mm

{\it Proof of Lemma} 6.3.
We first look for a lower bound for $t(c,\theta,x,s,u)$.
The following estimate is valid
$$
\eqalign{
\left|{s\theta\over\sqrt{1-s^2/c^2}}+{1\over s}\int_{-\infty}^u F(x+\tau\theta)d\tau\right|
\ge&\left|{s\theta\over\sqrt{1-s^2/c^2}}+{1\over s}\int_{-\infty}^{+\infty} F(x+\tau\theta)d\tau\right|\cr
&-\left|{1\over s}\int_u^{+\infty} F(x+\tau\theta)d\tau\right|.
}
\leqno(6.16)
$$
From (6.1) it follows that
$$
\left|{s\theta\over\sqrt{1-s^2/c^2}}+{1\over s}\int_{-\infty}^{+\infty} F(x+\tau\theta)d\tau\right|
\ge {s\over\sqrt{1-s^2/c^2}}-{2\beta_1\sqrt{d}\over \alpha(s/\sqrt{2})(1+|x|/\sqrt{2})^\alpha}.
\leqno(6.17)
$$
Using first (6.1b) and then $ s\ge z_2(c,d,\beta_1,\alpha,|x|)$ and (6.17) we obtain
$$
\leqalignno{ 
\left|{1\over s}\int_{u}^{+\infty} F(x+\tau\theta)d\tau\right|\le 
   &{\beta_1\sqrt{d}\over (s/\sqrt{2})\alpha(1+|x|/\sqrt{2})^\alpha}\cr
\le&{1\over 6}\left({s\over\sqrt{1-s^2/c^2}}-{2\beta_1\sqrt{d}\over \alpha(s/\sqrt{2})(1+|x|/\sqrt{2})^\alpha}\right)\cr
\le&{1\over 6}\left|{s\theta\over\sqrt{1-s^2/c^2}}+{1\over s}\int_{-\infty}^{+\infty} F(x+\tau\theta)d\tau\right|.&(6.18)
}
$$
From (6.16) and (6.18) it follows that
$$
\left|{s\theta\over\sqrt{1-s^2/c^2}}+{1\over s}\int_{-\infty}^u F(x+\tau\theta)d\tau\right|
\ge
{5\over 6}\left|{s\theta\over\sqrt{1-s^2/c^2}}+{1\over s}\int_{-\infty}^{+\infty} F(x+\tau\theta)d\tau\right|.
\leqno(6.19)
$$
Using (6.14) and (6.19) we obtain
$$
t(c,\theta,x,s,u)\ge {25\over 36} - 1=-{11\over 36}.
\leqno(6.20)
$$

Now we look for an upper bound for $t(c,\theta,x,s,u)$.
The right-hand side of (6.15) consists of a substraction of two fractions whose denominator is greater than $c^2$ and this implies
$$
\leqalignno{
|t(c,\theta,x,s,u)|\le&
   c^{-2}\left|-2V(x+u\theta)\sqrt{1-s^2/c^2}+{1-s^2/c^2\over s^2}\left|\int_u^{+\infty}F(x+\tau \theta)d\tau\right|^2\right.\cr
   &\left.-{2(1-s^2/c^2)\over s^2}\int_u^{+\infty}F(x+\tau \theta)d\tau.\int_{-\infty}^{+\infty}F(x+\tau \theta)d\tau\right|.&(6.21)
}
$$
Thus, using (1.2), (6.1), (4.7) and the fact $s\ge z_2(c,d,\beta_1,\alpha,|x|)$, we obtain
$$
\leqalignno{
|t(c,\theta,x,s,u)|\le& c^{-2}\sqrt{1-s^2/c^2}\left[2\beta_0(1+|x|/\sqrt{2}+u/\sqrt{2})^{-\alpha}\right.\cr
&+{\sqrt{1-z_2(c,d,\beta_1,\alpha,|x|)^2/c^2}
\over z_2(c,d,\beta_1,\alpha,|x|)^2}{d\beta_1^22\over\alpha^2(1+|x|/\sqrt{2}+u/\sqrt{2})^{2\alpha}}&(6.22)\cr
&\left.+{\sqrt{1-z_2(c,d,\beta_1,\alpha,|x|)^2/c^2}
\over z_2(c,d,\beta_1,\alpha,|x|)^2}{d\beta_1^28\over\alpha^2(1+|x|/\sqrt{2}+u/\sqrt{2})^\alpha(1+|x|/\sqrt{2})^\alpha}\right].
}
$$
Using (6.13), (6.20), the Taylor expansion of the map $]-1,+\infty[\mapsto \R$, $\delta\mapsto (1+\delta)^{-1/2}$ at $\delta=0$ and
(6.15), we obtain
$$
\leqalignno{
&\left|A(c,\theta,x,s,u)-1-{V(x+u\theta)\sqrt{1-s^2/c^2}\over c^2}\right|\cr
=&\left|{1\over2}t(c,\theta,x,s,u)+{3\over4}\int_0^1(1-w)(1+wt(c,\theta,x,s,u))^{-{5\over2}}dwt(c,\theta,x,s,u)^2
-{V(x+u\theta)\sqrt{1-s^2/c^2}\over c^2}\right|\cr
   \le&\left|{V(x+u\theta)\sqrt{1-s^2/c^2}\over c^2}\left[1-\left(1+{1-s^2/c^2\over c^2s^2}
   \left|\int_{-\infty}^{+\infty}F(x+\tau\theta)d\tau\right|^2\right)^{-1}\right]\right|\cr
     &+{1\over2}{{1-s^2/c^2\over s^2}\left|\int_u^{+\infty}F(x+\tau \theta)d\tau\right|^2
+{2(1-s^2/c^2)\over s^2}\int_u^{+\infty}|F(x+\tau \theta)|d\tau.\int_{-\infty}^{+\infty}|F(x+\tau \theta)|d\tau
\over \left(1+{(1-s^2/c^2)\over s^2c^2}\left|\int_{-\infty}^{+\infty}F(x+\tau \theta)d\tau\right|^2\right)c^2
}\cr
&\hskip 2cm+{3\over 8}({25\over36})^{-{5\over2}}t(c,\theta,x,s,u)^2.&(6.23)
}
$$
We use Lemma 6.2, conditions (1.2) and the fact that $s\ge z_2(c,d,\beta_1,\alpha,|x|)$ to estimate the first term of the right-hand 
side of the inequality (6.23).
In order to estimate the second term of the right-hand side of the inequality (6.23), we use the fact that the denominator is greater 
than $c^2$,  and we also use (6.1), (4.7), the fact that $s\ge z_2(c,d,\beta_1,\alpha,|x|)$.
We estimate the third term of the right-hand side of the inequality with (6.22).
Thus we obtain
$$
\left|A(c,\theta,x,s,u)-1-{V(x+u\theta)\sqrt{1-s^2/c^2}\over c^2}\right|
\le (1-s^2/c^2)h_{c,d,\beta_0,\beta_1,\alpha,|x|}(u),
$$
where
$$
\eqalign{
&h_{c,d,\beta_0,\beta_1,\alpha,|x|}(u)=\cr
   &{1\over c^2}(1+|x|/\sqrt{2}+u/\sqrt{2})^{-\alpha}\left\lbrace \beta_0k_{1,c,d,\beta_1,\alpha,|x|}
   \sqrt{1-z_2(c,d,\beta_1,\alpha,|x|)^2/c^2}\right.\cr
}
$$
$$
\eqalign{
&+{\beta_1^2d\over \alpha^2z_2(c,d,\beta_1,\alpha,|x|)^2}\left((1+|x|/\sqrt{2}+u/\sqrt{2})^{-\alpha}+4(1+|x|/\sqrt{2})^{-\alpha}\right)\cr
   &+{3\over2c^2}({25\over36})^{-{5\over2}}\left[\beta_0(1+|x|/\sqrt{2}+u/\sqrt{2})^{-\alpha/2}\right.\cr
   &+{d\beta_1^2\sqrt{1-z_2(c,d,\beta_1,\alpha,|x|)^2/c^2}\over z_2(c,d,\beta_1,\alpha,|x|)^2\alpha^2(1+|x|/\sqrt{2}+u/\sqrt{2})^{\alpha/2}}\cr
   &\times\left.\left.\left((1+|x|/\sqrt{2}+u/\sqrt{2})^{-\alpha}+4(1+|x|/\sqrt{2})^{-\alpha}\right)\right]^2\right\rbrace .\crcr
}
$$
Lemma 6.3 is proved.

\vskip 4mm

{\it Proof of Theorem} 3.2.

Let $(\theta, x)\in T\S^{d-1},\ \alpha,\ d,\ c,\ \beta_1,\ \beta_2,
\ s\in]0,c[,\ s\ge z_2(c,d,\beta_1,\alpha,|x|)$ be fixed.
We shall study the asymptotics of $l_{s\theta,x}(0)$ which is defined by formula (2.13b).

First we look for the asymptotics of 
$$
\int\limits_{-\infty}^0\left[g(\gamma(s\theta)+\int\limits_{-\infty}^\tau F(us\theta+x)du)-s\theta\right]d\tau.
$$
By changes of variables,  we obtain
$$
\leqalignno{
\int\limits_{-\infty}^0&\left[g(\gamma(s\theta)+\int\limits_{-\infty}^\tau F(us\theta+x)du)-s\theta\right]d\tau
=\cr
&\int\limits_{-\infty}^0\left[{{\theta\over \sqrt{1-s^2/c^2}}+{1\over s^2}\int\limits_{-\infty}^\tau F(u\theta+x)du
\over\sqrt{1+c^{-2}\left|{s\theta\over \sqrt{1-s^2/c^2}}+{1\over s}\int\limits_{-\infty}^\tau F(u\theta+x)du\right|^2}}-\theta\right]d\tau
.&(6.24)}
$$
Expanding the square of the norm in the denominator of the fraction under the integral in (6.24), the denominator becomes
$$
\leqalignno{\left(1+c^{-2}\left|{s\theta\over \sqrt{1-s^2/c^2}}+{1\over s}\int\limits_{-\infty}^\tau F(u\theta+x)du\right|^2\right)^{-1/2}
=&
\left(1+\delta_1(c,\theta,x,s,u)\right)^{-{1\over2}}\cr
&\times(1-s^2/c^2)^{1\over2},&(6.25)
}
$$
where $\delta_1$ is defined by formula (6.4).
We define
$$
\leqalignno{
\Lambda_1(\theta,x,s)=&\left|(1-s^2/c^2)^{-{1\over2}} 
\int\limits_{-\infty}^0\left[g(\gamma(s\theta)+\int\limits_{-\infty}^\tau F(us\theta+x)du)-s\theta\right]d\tau
\right.\cr
&\left.-c^{-2}\int_{-\infty}^0V(\tau\theta+x)d\tau\theta-s^{-2}\int_{-\infty}^0\!\int_{-\infty}^\tau F(u\theta+x)du d\tau
\right|.
&(6.26)
}
$$
From (6.26), (6.24) and (6.25), it follows that
$$
\leqalignno{
\Lambda_1(\theta,x,s)
\le&\int^0_{-\infty}\left|\left(1+\delta_1(c,\theta,x,s,u)\right)^{-{1\over2}}\right.
\cr
&\left.-1-c^{-2}V(\tau\theta+x)\sqrt{1-s^2/c^2}\theta
\right|\cr
&\times\left({1\over \sqrt{1-s^2/c^2}}+s^{-2}\int_{-\infty}^\tau |F(u\theta+x)|du\right)d\tau
\cr
&+\int^0_{-\infty}\left|(1+c^{-2}V(\tau\theta+x)\sqrt{1-s^2/c^2})
\left({\theta\over \sqrt{1-s^2/c^2}}+s^{-2}\int_{-\infty}^\tau F(u\theta+x)du\right)
\right.
\cr
&\left.-{\theta\over \sqrt{1-s^2/c^2}}-c^{-2}V(\tau\theta+x)\theta-s^{-2}\int_{-\infty}^\tau F(u\theta+x)du\right|d\tau
.&(6.27)\crcr
}
$$
We estimate the first integral of the right-hand side of (6.27) by the use of Lemma 6.1. Therefore expanding the first product under the
second integral of the form $\int_{-\infty}^0$ of the right-hand side of (6.27), we obtain
$$
\leqalignno{
\Lambda_1(\theta,x,s)\le&
\sqrt{1-s^2/c^2}\int_{-\infty}^0\left[\tilde{g}_{c,d,\beta_0,\beta_1,\alpha,|x|}(\tau)
\left(1+s^{-2}\sqrt{1-s^2/c^2}\int_{-\infty}^\tau |F(u\theta+x)|du\right)
\right.
\cr
&\left.+\left|{V(\tau\theta+x)\over s^2c^2}\int_{-\infty}^\tau F(u\theta+x)du\right|\right]d\tau
.
&(6.28)
}
$$
We define $|V|_{\infty}=\sup_{y\in\R^d}|V(y)|.$ Using (6.28), (1.2), (6.1a), (4.8) and the fact that 
$s\ge z_2(c,d,\beta_1,\alpha,|x|),$ we obtain
$$
\leqalignno{
\Lambda_1(\theta,x,s)\le&
\sqrt{1-s^2/c^2}\left[\left(1+{\sqrt{1-z_2(c,d,\beta_1,\alpha,|x|)^2/c^2}\over
z_2(c,d,\beta_1,\alpha,|x|)^2}{\beta_1\sqrt{d}\sqrt{2}\over \alpha (1+|x|/\sqrt{2})^{\alpha}}\right)
\int_{-\infty}^0\tilde{g}_{c,d,\beta_0,\beta_1,\alpha,|x|}(\tau)d\tau\right.
\cr
&\left.+{\beta_1\sqrt{d}2|V|_{\infty}\over{z_2(c,d,\beta_1,\alpha,|x|)^2c^2\alpha(\alpha-1)(1+|x|/\sqrt{2})^{\alpha-1}}}
\right]
.&(6.29)
\crcr
}
$$
\vskip 4mm

Now we look for the asymptotics of 
$$
\int\limits_0^{+\infty}\left[g(\gamma(s\theta)+\int\limits_{-\infty}^\tau F(us\theta+x)du)-
		  g(\gamma(s\theta)+\int\limits_{-\infty}^{+\infty} F(us\theta+x)du)\right]d\tau.
$$
By changes of variables, we obtain
$$
\leqalignno{
&\int\limits_0^{+\infty}\left[g(\gamma(s\theta)+\int\limits_{-\infty}^\tau F(us\theta+x)du)-
		  g(\gamma(s\theta)+\int\limits_{-\infty}^{+\infty} F(us\theta+x)du)\right]d\tau\cr
=&
\int\limits_0^{+\infty}\left[{{\theta\over \sqrt{1-s^2/c^2}}+{1\over s^2}\int\limits_{-\infty}^\tau F(u\theta+x)du
\over\sqrt{1+c^{-2}\left|{s\theta\over \sqrt{1-s^2/c^2}}+{1\over s}\int\limits_{-\infty}^\tau F(u\theta+x)du\right|^2}}
\right.\cr
&\hskip 3cm \left.-{{\theta\over \sqrt{1-s^2/c^2}}+{1\over s^2}\int\limits_{-\infty}^{+\infty}F(u\theta+x)du
\over\sqrt{1+c^{-2}\left|{s\theta\over \sqrt{1-s^2/c^2}}+{1\over s}\int\limits_{-\infty}^{+\infty} 
F(u\theta+x)du\right|^2}}\right]d\tau
.&(6.30)}
$$
First we study the denominator of the first fraction under the integral of (6.30). From (6.14) and (6.13) it follows that
$$
\left(1+c^{-2}\left|{s\theta\over \sqrt{1-s^2/c^2}}+{1\over s}\int\limits_{-\infty}^\tau F(u\theta+x)du\right|^2\right)^{-{1\over2}}
$$
$$
=\left(1+c^{-2}\left|{s\theta\over \sqrt{1-s^2/c^2}}+{1\over s}\int\limits_{-\infty}^{+\infty}F(u\theta+x)du\right|^2\right)^{-{1\over 2}}
A(c,\theta,x,s,\tau).
\leqno(6.31)
$$
We define
$$
\leqalignno{
\Lambda_2(c,\theta,x,s)=&\left|(1-s^2/c^2)^{-{1\over2}} 
\int\limits^{+\infty}_0\left[g(\gamma(s\theta)+\int\limits^{+\infty}_\tau F(us\theta+x)du)-g(\gamma(s\theta)
+\int\limits^{+\infty}_{-\infty} F(us\theta+x)du)\right]d\tau
\right.\cr
&\left.-c^{-2}\int^{+\infty}_0V(\tau\theta+x)d\tau\theta+s^{-2}\int^{+\infty}_0\!\int^{+\infty}_\tau F(u\theta+x)du d\tau
\right|.
&(6.32)
}
$$
From (6.30), (6.31) and (6.32) it follows that
$$
\Lambda_2(c,\theta,x,s)\le \Lambda_{2,1}(c,\theta,x,s)+\Lambda_{2,2}(c,\theta,x,s),
\leqno(6.33)
$$
where
$$
\leqalignno{
\Lambda_{2,1}(c,\theta,x,s)
=&\int_0^{+\infty}\left|(1-s^2/c^2)^{-{1\over2}}
\left(1+{1\over c^2}\left|{s\theta\over \sqrt{1-s^2/c^2}}
+{1\over s}\int\limits_{-\infty}^{+\infty}F(u\theta+x)du\right|^2\right)^{-{1\over 2}}
\right.\cr
&\left.\times\left(A(c,\theta,x,s,\tau)-1-V(\tau\theta+x){\sqrt{1-{s^2\over c^2}}\over c^2}\right)
\left({\theta\over \sqrt{1-{s^2\over c^2}}}+{1\over s^2}\int\limits_{-\infty}^\tau\!\!\! F(u\theta+x)du\right)\right|d\tau,&(6.34\rm a)
\cr
\Lambda_{2,2}(c,\theta,x,s)
=&\int_0^{+\infty}\left|(1-s^2/c^2)^{-{1\over2}}
\left(1+{1\over c^2}\left|{s\theta\over \sqrt{1-s^2/c^2}}
+{1\over s}\int\limits_{-\infty}^{+\infty}F(u\theta+x)du\right|^2\right)^{-{1\over 2}}
\right.
\cr
&\times\left.\left[\left(1+V(\tau\theta+x){\sqrt{1-s^2/c^2}\over c^2}\right)
\left({\theta\over \sqrt{1-s^2/c^2}}+{1\over s^2}\int\limits_{-\infty}^\tau F(u\theta+x)du\right)
\right.\right.\cr
&\left.\left.-\left({\theta\over \sqrt{1-s^2/c^2}}+{1\over s^2}\int\limits_{-\infty}^{+\infty} F(u\theta+x)du\right)
\right]
-{V(\tau\theta+x)\over c^2}\theta+{1\over s^2}\int\limits_{\tau}^{+\infty} F(u\theta+x)du\right|d\tau.&(6.34\rm b)
}
$$
Let us estimate $\Lambda_{2,1}(c,\theta,x,s).$

From Lemma 6.3 and (6.34a) it follows that
$$
\leqalignno{\Lambda_{2,1}(c,\theta,x,s)
\le&\sqrt{1-s^2/c^2} \left(1+{1\over c^2}\left|{s\theta\over \sqrt{1-s^2/c^2}}
+{1\over s}\int\limits_{-\infty}^{+\infty}F(u\theta+x)du\right|^2\right)^{-{1\over 2}}\cr
&\times\left({1\over\sqrt{1-s^2/c^2}}+{1 \over s^2}\int\limits_{-\infty}^{+\infty}|F(u\theta+x)|du
\right)
\times \int_0^{+\infty}h_{c,d,\beta_0,\beta_1,\alpha,|x|}(\tau)d\tau.
&(6.35)
}
$$
In addition, expanding the square of the norm, we obtain
$$
\leqalignno{
\left(1+{1\over c^2}\left|{s\theta\over \sqrt{1-s^2/c^2}}
+{1\over s}\int\limits_{-\infty}^{+\infty}F(u\theta+x)du\right|^2\right)^{-{1\over 2}}
=&\sqrt{1-s^2/c^2}\cr
&\times \left(1
+{1-s^2/c^2\over s^2c^2}\left|\int\limits_{-\infty}^{+\infty}F(u\theta+x)du\right|^2\right)^{-{1\over 2}}
&(6.36\rm a)\cr
\le&\sqrt{1-s^2/c^2}.&(6.36\rm b)
}
$$
Using (6.35), (6.36), (4.1), (6.2), (4.7) and the fact that 
$s\ge z_2(c,d,\beta_1,\alpha,|x|),$ we obtain
$$
\leqalignno{\Lambda_{2,1}(c,\theta,x,s)\le&
\sqrt{1-s^2/c^2}\left(1+{\beta_1 \sqrt{d}\sqrt{1-z_2(c,d,\beta_1,\alpha,|x|)^2/c^2}2\sqrt{2}\over z_2(c,d,\beta_1,\alpha,|x|)^2\alpha(1+|x|/\sqrt{2})^\alpha}
\right)\cr
&\times \int_0^{+\infty}h_{c,d,\beta_0,\beta_1,\alpha,|x|}(\tau)d\tau.&(6.37)
}
$$

Let us estimate $\Lambda_{2,2}(c,\theta,x,s).$

From (6.34b) and (6.36a) it follows that
$$
\leqalignno{
\Lambda_{2,2}(c,\theta,x,s)
\le&\int_0^{+\infty}
\left|\left(1+{1-s^2/c^2\over s^2c^2}
\left|\int\limits_{-\infty}^{+\infty}F(u\theta+x)du\right|^2\right)^{-{1\over 2}}
-1\right|
\cr
&\times\left|-{1\over s^2}\int\limits^{+\infty}_\tau F(u\theta+x)du+{V(\tau\theta+x)\over c^2}\theta
+{\sqrt{1-s^2/c^2}\over s^2c^2}V(\tau\theta+x)\int\limits_{-\infty}^\tau F(u\theta+x)du\right|d\tau
\cr
&+\int_0^{+\infty}\left|{\sqrt{1-s^2/c^2}\over s^2c^2}V(\tau\theta+x)\int\limits_{-\infty}^\tau F(u\theta+x)du\right|d\tau
.&(6.38)
\cr
}
$$
Thus using Lemma 6.2, conditions (1.2), (4.1), (6.2), (4.6), (4.7) and the fact that 
$s\ge z_2(c,d,\beta_1,\alpha,|x|),$ it follows that
$$
\leqalignno{
\Lambda_{2,2}(c,\theta,x,s)
\le&
\sqrt{1-s^2/c^2}\left[{k_{1/2,c,d,\beta_1,\alpha,|x|}\sqrt{1-z_2(c,d,\beta_1,\alpha,|x|)^2/c^2}\over (\alpha-1)(1+|x|/\sqrt{2})^{\alpha-1}}
\left({\sqrt{d}\beta_12\over z_2(c,d,\beta_1,\alpha,|x|)^2 \alpha}\right.\right.
\cr
&\left.\left.+{\beta_0\sqrt{2}\over c^2}
+{\sqrt{d}\beta_0\beta_14\sqrt{1-z_2(c,d,\beta_1,\alpha,|x|)^2/c^2}\over z_2(c,d,\beta_1,\alpha,|x|)^2 c^2\alpha(1+|x|/\sqrt{2})^{\alpha}}
\right)\right.
\cr
&\left.+{\sqrt{d}\beta_0\beta_14\over z_2(c,d,\beta_1,\alpha,|x|)^2c^2\alpha(\alpha-1)(1+|x|/\sqrt{2})^{2\alpha-1}}\right].
&(6.39) 
}
$$

From (2.13b), (6.26), (6.29), (6.32), (6.34), (6.35) and (6.39) it follows that there exists $C_{c,d,\beta_0,\beta_1,\alpha,|x|}$ such that
$$
\leqalignno{
\left|{l_{s\theta,x}(0)\over \sqrt{1-s^2/c^2}}
-\right.&{1\over c^2}PV(\theta,x)\theta+{1\over s^2}\left.\int_0^{+\infty}\!\int_{\tau}^{+\infty}F(x+u\theta)du d\tau
-{1\over s^2}\int^0_{-\infty}\!\int^{\tau}_{-\infty}F(x+u\theta)du d\tau\right|\cr
\le& \Lambda_1(c,\theta,x,s)+\Lambda_2(c,\theta,x,s)\cr
\le&C_{c,d,\beta_0,\beta_1,\alpha,|x|}\sqrt{1-s^2/c^2}
.&(6.40)
}
$$
The estimate (3.9) follows from (6.40). 

Theorem 3.2 is proved.

\vskip 4mm

{\it Proof of Proposition} 1.1.
The item 1 follows immediately from
$$
{{\rm d}\over {\rm d}t}V(t\theta+x)=\nabla V(t\theta+x)\theta,\ {\rm for\ all\ } (\theta,x)\in T\S^{d-1},\ t\in \R.
$$
Proof of the item 2.
Take 
$$
V(x)={x_1\over (1+|x|^2)^{\beta}}, {\rm\ for\ } x=(x_1,\ldots,x_d)\in\R^d, \beta>1.
$$ 
and take $(\theta,x)\in T\S^{d-1}$. By a straightforward calculation and using $\theta x=0$,
we obtain
$$
\eqalign{
\left(\int_{-\infty}^0\!\int_{-\infty}^{\tau} F(s\theta+x)dsd\tau\right.-&\left.\int_0^{+\infty}\!\int^{+\infty}_{\tau} 
F(s\theta+x)dsd\tau+PV(\theta,x)\theta\right)\star x\cr
=&-4\beta \theta_1|x|^2\int_0^{+\infty}\!\int^{+\infty}_{\tau}{s\over (1+s^2+|x|^2)^{\beta+1}}dsd\tau\cr
\neq&\ 0 {\rm\ if\ and \ only\ if\ }x\neq0 {\rm\ and\ }\theta_1\neq0,
\cr 
}
$$
where $\star$ denotes the scalar product.

Proof of the item 3.
Let $V$ be a spherical symmetric potential (i.e. $V$ takes the form $m(|x|)$) that satisfies the conditions (1.2)
(e.g. $V(x)=(1+|x|^2)^{-\beta}$ where $\beta>{1\over 2}$). Then $m\in C^1(]0,+\infty[, \R)$ and $\nabla V(x)=m'(|x|){x\over |x|}$. 
Let $(\theta,x)\in T\S^{d-1}$ and let $\theta^{\bot}$ be an orthogonal vector to $\theta$. A straighforward calculation gives
$$
F(s\theta+x)\theta^\bot=m'(\sqrt{s^2+|x|^2}){x \star \theta^{\bot} \over \sqrt{s^2+|x|^2}}
$$
for any $s\in \R$. Hence
$$
\left(\int_{-\infty}^0\!\int_{-\infty}^{\tau} F(s\theta+x)dsd\tau-\int_0^{+\infty}\!\int^{+\infty}_{\tau} F(s\theta+x)dsd\tau
+PV(\theta,x)\theta\right)\star\theta^{\bot} 
=0.\leqno(6.41)
$$
The item 3 follows from the item 1 and formula (6.41). 

Proposition 1.1 is proved.

\vskip 1cm

\noindent{\bf Acknowledgement.} This work was fulfilled in the framework of Ph. D. thesis researchs under the direction of R.G. Novikov.

\vskip 1cm

\centerline{\bf References}
\vskip 2 mm

\item{[E]\hfill} Einstein, A.: \" Uber das Relativit\" atsprinzip und die aus demselben gezogenen Folgerungen. Jahrbuch der Radioaktivit\" at
und Elektronik {\bf 4},  411-462 (1907)
\item{[GGG]} Gel'fand, I.M., Gindikin, S.G., Graev, M.I.: Integral geometry
in affine and projective spaces. Itogi Nauki i Tekhniki, Sovr. Prob. Mat.
{\bf 16}, 53-226 (1980) (Russian)
\item{[GN]\hfill} Gerver, M.L., Nadirashvili, N. S.:  Inverse problem of mechanics at high energies. Comput. Seismology {\bf 15}, 118-125 (1983)
(Russian)
\item{[LL1]} Landau, L.D., Lifschitz, E.M.: {\it Mechanics}. Oxford: Pergamon Press, 1960
\item{[LL2]} Landau, L.D., Lifschitz, E.M.: {\it The Classical Theory of Fields}. New York: Pergamon Press, 1971
\item{[Na]\hfill} Natterer, F.: {\it The Mathematics of Computerized Tomography}. Stuttgart: Teubner and  
Chichester: Wiley, 1986
\item{[No]\hfill} Novikov, R.G.: Small angle scattering and X-ray transform
in classical mechanics. Ark. Mat. {\bf 37},  141-169 (1999)
\item{[R]\hfill} Radon, J.: \" Uber die Bestimmung von Funktionen durch ihre Integralwerte l\" angs 
 gewisser Mannigfaltigkeiten. Ber. Verh. S\" achs. Akad. Wiss. Leipzig, Math.-Nat. K1 
{\bf 69}, 262-277 (1917)
\item{[S]\hfill} Simon, B.: Wave operators for classical particle scattering. Comm. Math. Phys. {\bf 23}, 37-48 (1971)
\item{[Y]\hfill} Yajima, K.: Classical scattering for relativistic particles. J. Fac. Sci., Univ. Tokyo, 
Sect. I A {\bf 29}, 599-611 (1982)

\end